\documentclass[rsc,preprint,amsmath,amssymb,amsfonts,floatfix,superscriptaddress]{revtex4}

\usepackage{graphicx}
\usepackage{lscape}
\usepackage{float}
\usepackage{psfrag}
\usepackage{algpseudocode}
\usepackage{appendix}
\usepackage{caption}
\usepackage{amsmath}
\usepackage{subcaption}
\usepackage{overpic}
\usepackage{xr}
\makeatletter
\newcommand*{\addFileDependency}[1]{
  \typeout{(#1)}
  \@addtofilelist{#1}
  \IfFileExists{#1}{}{\typeout{No file #1.}}
}
\makeatother

\newcommand*{\myexternaldocument}[1]{%
    \externaldocument{#1}%
    \addFileDependency{#1.tex}%
    \addFileDependency{#1.aux}%
}
\myexternaldocument{si}

\begin{document}

\title{Resolving Competing Conical Intersection Pathways: Time-Resolved X-ray Absorption Spectroscopy of trans-1,3-Butadiene}

\author{Issaka Seidu}
\affiliation{National Research Council Canada, 100 Sussex Drive, Ottawa, Ontario K1A 0R6, Canada}

\author{Simon P. Neville}
\affiliation{National Research Council Canada, 100 Sussex Drive, Ottawa, Ontario K1A 0R6, Canada}

\author{Ryan J. MacDonell$^*$}
\affiliation{Department of Chemistry and Biomolecular Sciences, University of Ottawa, 10 Marie Curie, Ottawa, Ontario, K1N 6N5, Canada}
\thanks{present address: School of Chemistry, University of Sydney, NSW 2006, Australia}

\author{Michael S. Schuurman}
\affiliation{National Research Council Canada, 100 Sussex Drive, Ottawa, Ontario K1A 0R6, Canada}
\affiliation{Department of Chemistry and Biomolecular Sciences, University of Ottawa, 10 Marie Curie, Ottawa, Ontario, K1N 6N5, Canada}

\begin{abstract}
  Time-resolved X-ray absorption spectroscopy is emerging as a uniquely powerful tool to probe coupled electronic-nuclear dynamics in photo-excited molecules. Theoretical studies to date have established that time-resolved X-ray absorption spectroscopy is an atom-specific probe of excited-state wave packet passage through a seam of conical intersections (CIs). However, in many molecular systems, there are competing dynamical pathways involving CIs of different electronic and nuclear character. Discerning these pathways remains an important challenge.  Here, we demonstrate that time-resolved X-ray absorption spectroscopy (TRXAS) has the potential to resolve competing channels in excited-state non-adiabatic dynamics. Using the example of 1,3-butadiene, we show how TRXAS discerns the different 
  electronic structures associated with passage through multiple conical intersections. Trans 1,3-butadiene exhibits a branching between \textit{polarized} and \textit{radicaloid} pathways associated with ethylenic ``twisted-pyramidalized'' and excited-state cis-trans isomerization dynamics, respectively.  The differing electronic structures along these pathways give rise to different XAS signals, indicating the possibility of resolving them. Furthermore, this indicates that XAS, and other core-level spectroscopic techniques, offer the appealing prospect of directly probing the effects of selective chemical substitution and its ability to affect chemical control over excited-state molecular dynamics. 
\end{abstract}

\maketitle

\section{Introduction}
Time-resolved spectroscopy has been extremely successful in the
experimental interrogation of conical intersection (CI) mediated
dynamics in molecular photo-physical
processes\cite{Blanchet_Nature_1999,Stolow_AnnuRevPhysChem_2003,Stolow_ChemRev_2004,Hofmann_ChemPhysLett_2001,Schuurman_AnnuRevPhysChem_2018,Voll_JPhotochemPhotobiol_2007,Geneaux_2019,Boguslavskiy_JChemPhys_2018}. 
However, it is often the case that the time-evolution of specific spectroscopic
observables display non-intuitive ``mappings'' to the
time-evolution of the underlying coupled nuclear-electronic
dynamics. The difficulty of uniquely assigning observable spectroscopic
quantities to changes in the electronic and nuclear character of an excited-state
wave packet has been previously addressed via the first-principles simulation
of these time-resolved spectra\cite{Hudock_2007,Glover_JChemPhys_2018, MacDonell_JPhysChemA_2019,Tsuru_StructDyn_2021,Golubev_PhysRevLett_2021,Northey_PhysChemChemPhys_2020}. The rationale for this approach is that an accurate time-resolved spectral simulation will engender confidence in the underlying molecular dynamics calculations used to generate the computed spectrum. It is from these dynamics simulations that (detailed) mechanistic elements may be extracted.

For systems that undergo CI-mediated non-adiabatic transitions, a key
mechanistic detail that is often extracted from excited-state dynamics simulations is the identity of the CI seam(s) involved (i.e. the nuclear and electronic structure) in the
electronic relaxation process. This detailed information can be difficult to discern from a spectroscopic
signal alone. However, it has been recently demonstrated that
core-level spectroscopy will offer a uniquely powerful means to track the
change in electronic structure in the region of strong non-adiabatic
coupling surrounding a CI seam\cite{Golubev_PhysRevLett_2021,Northey_PhysChemChemPhys_2020,Neville_PhysRevLett_2018,Zinchenko_Science_2021},
and may thus encode the approach of a wave packet to it. In many molecular systems, though, the electronic relaxation of an excited-state wave packet is not dominated by a single localized region of the CI seam.
In fact, there may exist competition between multiple conical intersections and/or multiple seams, with a branching
ratio determined by both nuclear (inertial) and electronic effects inherent in the prepared wave packet\cite{Schuurman_AnnuRevPhysChem_2018,Wu_2011,MacDonell_2016,MacDonell_ChemPhys_2018,MacDonell_JPhysChemA_2019,Herperger_JChemPhys_2020}.

The question posed in this work is this: to what extent is X-ray absorption
spectroscopy sensitive to the different electronic structures associated with multiple, ``competing'', conical intersection-mediated dynamical pathways? 
We will attempt to answer this question via simulation using as a paradigm example, the non-adiabatic dynamics of
1,3-butadiene (1,3-BD) following excitation to the bright $\pi\pi^{*}$
state. In the ensuing multi-state dynamics, there exist (at least) two important nuclear configurations that are associated with internal conversion to the ground electronic states: the so-called twisted-pyramidalized (Tw-Py) and
trans-cis isomerization (transoid) conical intersections\cite{Levine_JPhysChemA_2009,Glover_JChemPhys_2018,MacDonell_JPhysChemA_2019}. For reference, the Tw-Py and transoid minimum energy CI (MECI) geometries are
shown in Fig.~\ref{fig:meci_geom}. The Tw-Py MECI is analogous to the
twisted-pyramidalized intersection motif in ethylene\cite{BenNum_ChemPhysLett_1998,Mori_JPhysChemA_2012}, where the electronic structure of the $S_{1}$ state is
characterized by a separation of charge between the terminal and medial carbon atoms. There is an additional CI observed in simulation as a minor channel, in which the pyramidalization occurs at a medial (rather than terminal) C atom, which is denoted ``mTw-Py" in Fig.~\ref{fig:meci_geom}(c).
The transoid structure is realized via twisting about the C-C-C-C backbone (dihedral $\angle_{Tw-B}$ = 56.151$^{\circ}$) in concert with a slight twisting of the terminal methylene groups\cite{Levine_JPhysChemA_2009}. This twisting is accompanied by a mixing of the excited $\pi^{*2}$ and ground ($\pi^{*2} - S_0$) electronic states. These transoid (and cisoid, together
termed ``kinked-diene'') structures have a tetraradical character and thus the lack of
electronic polarization that is observed in the former pathway\cite{MacDonell_JPhysChemA_2019}.
Numerous previous works found that these pathways are \textit{both} involved in the electronic de-excitation, and that the branching ratio between the two
could be modified via chemical substitution\cite{MacDonell_JPhysChemA_2019}, which affects both the relative ``inertia'' associated with the twisting as well as the relative energies of the different CI seam regions.\cite{MacDonell_ChemPhys_2018,MacDonell_JPhysChemA_2019} 

Previous joint theoretical-experimental efforts using valence time-resolved photo-electron spectroscopy (TRPES)\cite{Boguslavskiy_JChemPhys_2018,Glover_JChemPhys_2018,MacDonell_JChemPhys_2020} have provided incredible insight into: (\textit{i}) the time-scale for electronic relaxation back to the ground state, (\textit{ii}) how the initial nuclear motion involving the torsional coordinate can be encoded in the TRPES, and importantly for this work, (\textit{iii}) how \textit{ab initio} determinations of the TRPES based on non-adiabatic dynamics simulations are able to successfully capture many of the key features of the experimental spectra. What is more difficult to discern from these previous studies, however, are the spectral fingerprints of the CIs themselves. Any signal that simulation predict arises uniquely from the CI region is itself embedded in broad ionization features associated more generally with the excited electronic state(s).  Rather, the nuclear and electronic character of these pathways are discerned indirectly via simulation alone, where the degree of fidelity between computed and measured (global) spectrum strongly informs how convincing such assignments may be. Clearly, a more ``ideal'' observable would be able to isolate those structures that simulation predicts are so central to the electronic decay process.

In the present work, we use a combination of \textit{ab initio} 
multiple spawning (AIMS) dynamics simulations\cite{BenNun_JPhysChemA_2000,Martinez_AccChemRes_2006} and X-ray absorption cross-section calculations,  to simulate the time-resolved X-ray absorption spectrum (TRXAS) of trans 1,3-BD following excitation to the bright $(\pi\pi^{*})$ states. In the following, the ability of TRXAS to resolve \textit{both} the nuclear and electronic characters of the seam regions accessed by the excited-state wave packet is interrogated via simulation.

\section{Theoretical methods}

\subsection{Dynamics simulations}
The excited-state dynamics of 1,3-BD was simulated
using the AIMS method\cite{BenNun_JPhysChemA_2000,Martinez_AccChemRes_2006}, which in this study
involves invoking the independent first generation (IFG) approximation and employing a zeroth-order saddle-point approximation for the evaluation of the requisite Hamiltonian matrix elements over trajectory basis functions\cite{churchod_2018}. Here, the total molecular wave function is given by:

\begin{equation}
  \Psi \left( \boldsymbol{R}, \boldsymbol{r}, t \right) =
  \sum_{I=1}^{N_{s}} \chi_{I}(\boldsymbol{R}, t) \psi_{I}\left(
  \boldsymbol{r}; \boldsymbol{R} \right),
\end{equation}

\noindent
where $\boldsymbol{r}$ and $\boldsymbol{R}$ denote the electronic and
nuclear coordinates. Here, $\psi_{I}$ is the $I$th adiabatic electronic
state, and $\chi_{I}$ is the corresponding nuclear wave function
expressed in terms of frozen Gaussian basis functions, $g_{j}^{I}$:

\begin{equation}
  \chi_{I} \left(\boldsymbol{R}, t \right) = \sum_{j=1}^{N_{I}(t)}
  C_{j}^{I}(t) g_{j}^{I} \left( \boldsymbol{R};
  \boldsymbol{\alpha}_{j}^{I}, \bar{\boldsymbol{R}}_{j}^{I}(t),
  \bar{\boldsymbol{P}}_{j}^{I}(t), \gamma_{j}^{I}(t)\right).
\end{equation}

\noindent
The time-dependent Gaussian positions and momenta,
$\bar{\boldsymbol{R}}_{j}^{I}(t)$ and
$\bar{\boldsymbol{P}}_{j}^{I}(t)$, respectively, evolve
according to classical equations of motion. The phases
$\gamma_{j}^{I}(t)$ are propagated according to semi-classical
equations, and the Gaussian widths $\boldsymbol{\alpha}_{j}^{I}$ are
time-independent. The expansion coefficients, $C_j^I(t)$, are
determined variationally by solving the time-dependent Schr{\"o}dinger
equation. During an AIMS calculation, the number of Gaussian basis
functions, $N_I$, is increased in regions of strong non-adiabatic
coupling between electronic states in a process referred to as
spawning, providing a description of internal conversion processes\cite{Yang_JChemPhys_2009}.

\subsection{Simulation of time-resolved X-ray absorption spectra}
The TRXAS was simulated as the incoherent sum of the X-ray absorption
spectra computed at the centroid of each Gaussian basis function of
the AIMS simulation, with a weighting factor given by the squared
modulus of the corresponding expansion coefficient. That is, the TRXAS
was approximated as

\begin{equation}
  \sigma(E,t) = \sum_{I=1}^{N_{s}} \sum_{j=1}^{N_{j}^{I}} \left|
  C_{j}^{I}(t) \right|^{2} \sigma_{I} \left(E;
  \bar{\boldsymbol{R}}_{j}^{I}(t) \right),
  \label{eq:conv}
\end{equation}

\noindent
where $\sigma_{I}(E; \boldsymbol{R})$ denotes the X-ray absorption
spectrum of the $I$th electronic state computed at the nuclear
geometry $\boldsymbol{R}$. This approach is completely analogous to 
that employed in the simulation of valence spectroscopies\cite{Hudock_2007}. Here, though, we employ core-excitation oscillator strengths as probe-induced observable instead of, for example,
the norms of Dyson orbitals as approximate ionization cross-sections in the case of
first-principles TRPES simulations. 

For the calculation of the excited-state X-ray absorption
cross-sections, the core-valence separated combined density functional
theory and multi-configurational configuration interaction
(CVS-DFT/MRCI) method was used\cite{Grimme_DFTCIS_ChemPhysLett_1996,Grimme_DFTMRCI_JCP_1999,Heil_DFTMRCI_MolPhys_2016,Heil_DFTMRCI_JCP_2017,Lyskov_DFTMRCI_JCP_2016,Kleinschmidt_DFTMRCI_JCompChem_2002,Marian_2019,Seidu_JChemPhys_2019}. 
This method has a key advantage of being able to correctly describe the
types of multi-reference and doubly-excited core-excited states that
often contribute to the X-ray absorption spectra of
singly-valence-excited states\cite{Seidu_JChemPhys_2019}. The ability to accurately handle states of these types is especially important in this application due to the complex nature of valence electronic states encountered. This is a key advantage of using DFT/MRCI compared to single-reference methods\cite{Marian_2019}.

The present simulations consider the isotropic axis distribution TRXAS (see Fig.~\ref{fig:trxas_iso}) as well as the relative polarization of the pump and probe pulses, and we display the results for both the parallel and perpendicular pump/probe relative polarization in the SI, Figs.~S2 and S3. Specifically, these polarization-dependent spectra were simulated assuming that the pump pulse generates a $\cos(\theta)^2$ distribution in the excited-state population, where $\theta$ is the angle between pump polarization and the transition dipole moment for the $S_0 \rightarrow S_1$ transition. Practically, this done by weighting the oscillator strength of the core-transition by the $\cos(\theta)^2$ axis distribution function, where $\theta$ is the angle between the pump and probe transition dipole moments. It was observed that this polarization resulted in the majority of the signal intensity in the perpendicular TRXAS. 

\subsection{Computational details}
The electronic energies, gradients and non-adiabatic couplings required for the AIMS simulations were computed at the multi-reference configuration interaction with single excitations (MR-CIS) level of theory\cite{Lischka_WIREs_2011,Lischka_COLUMBUS_2015}. An active space of 4 electrons in 4 orbital was used, comprising of HOMO and HOMO-1 $\pi$ orbitals, and
the LUMO and LUMO+1 $\pi^*$ orbitals. The 6-31G$^{*}$ basis set was used for the molecule. In the AIMS simulations, three electronic states were included, correlating with the $S_{0}$, $S_{1}(\pi^{*2})$ and $S_{2}(\pi\pi^{*})$ states at the FC point. A total of 39 initial conditions were initiated on the bright state, sampled from the ground-state Wigner distribution. For more information on the dynamics calculation, see Ref.~\citenum{MacDonell_JPhysChemA_2019}. The nuclear densities in internal coordinates were obtained via Monte-Carlo integration of Cartesian trajectory
basis functions using an approach analogous to that described in Ref.~\cite{Coe_2007}

In the DFT/MRCI calculations of the X-ray absorption cross-sections, the redesigned Hamiltonian of Heil \textit{et al.} was used\cite{Lyskov_DFTMRCI_JCP_2016}. The Turbomole set of programs\cite{turbomole_v6.1} and the Dalton software package\cite{daltonpaper} were used for the calculation of the KS-DFT orbitals in these calculations. Here, the BH-LYP functional\cite{Becke_B3LYP_PhysRevA_1988,Becke_B3LYP_JCP_1993,Becke_JChemPhys_1993} and aug-cc-pVDZ basis set were used exclusively.

We note that the level of electronic structure used in the dynamics and X-ray absorption cross-section calculations are different. As a consequence, the mapping between the sets of DFT/MRCI and MR-CIS
valence-state adiabatic electronic wave functions had to be computed at each geometry $\bar{\boldsymbol{R}}_{j}^{I}$ arising from the AIMS simulations. This was achieved by calculating the overlap of the MR-CIS electronic state associated with each trajectory and a corresponding set of  DFT/MRCI electronic wave functions. For 
 $>99.9\%$ of all geometries and $100\%$ of excited-state geometries, 
there was found to exist a DFT/MRCI state with an overlap $>0.90$ with each MR-CIS state of interest, enabling an unambiguous mapping between the different adiabatic states.

\section{Results and Discussion}

\subsection{Excited-State Dynamics of 1,3-BD}

We begin with an overview of the AIMS dynamics simulations of 1,3-BD. These simulations utilized the same level of electronic structure as in Ref.~\citenum{MacDonell_JPhysChemA_2019} and the reader is directed to that work for a more detailed discussion of the excited-state dynamics. Here, we will only briefly summarize those results, while performing a number of new analyses to highlight those aspects of the dynamics that pertain directly to our discussion of the TRXAS.

\begin{figure}[hbt!]
    \centering
    \includegraphics[width=.7\textwidth]{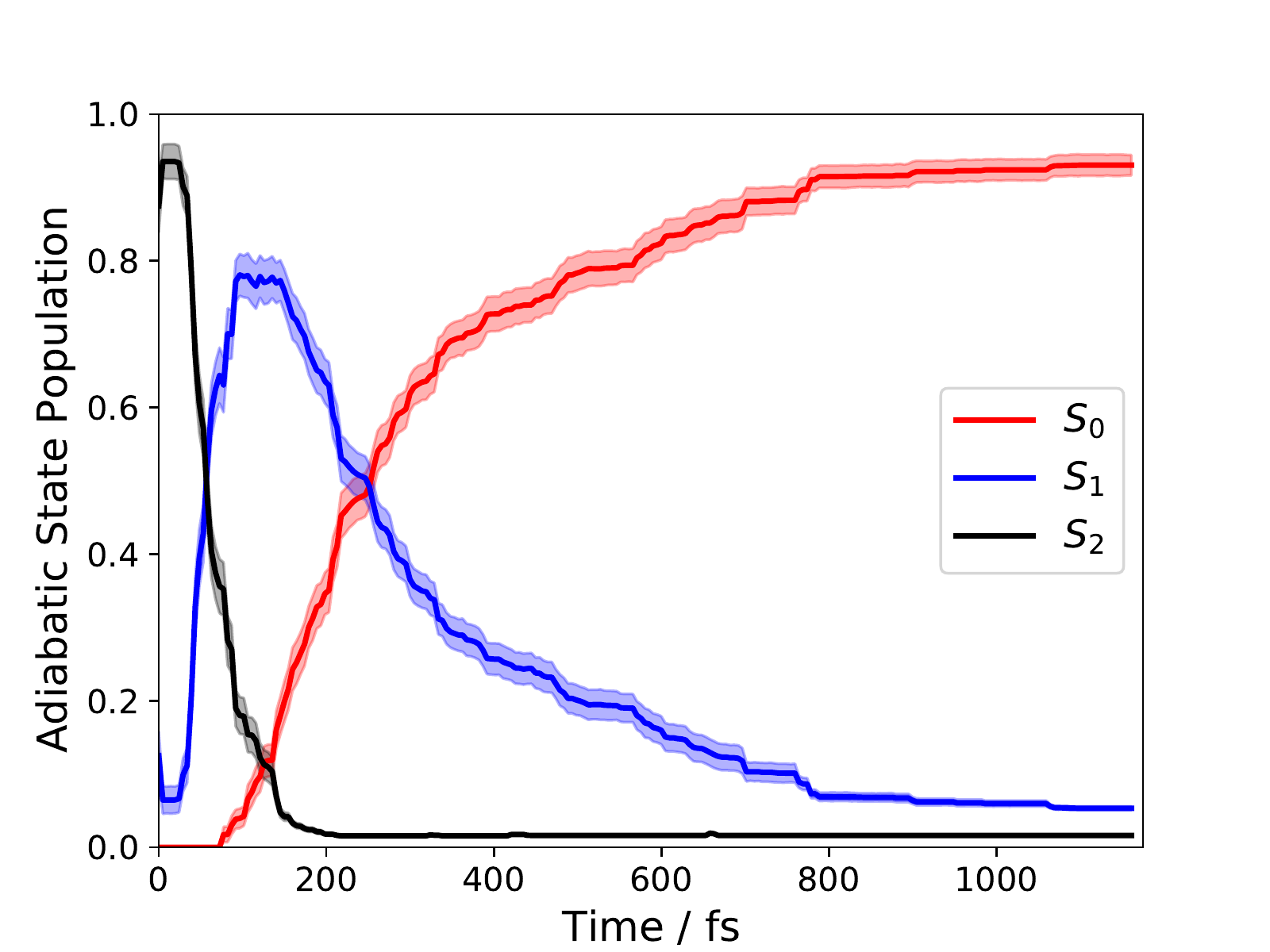}
    \caption{Adiabatic state populations calculated using AIM following vertical excitation into $S_1$ for 1,3-BD.}
    \label{fig:adpop}
\end{figure}

\begin{figure}[hbt!]
    \centering
    \includegraphics[width=.5\textwidth]{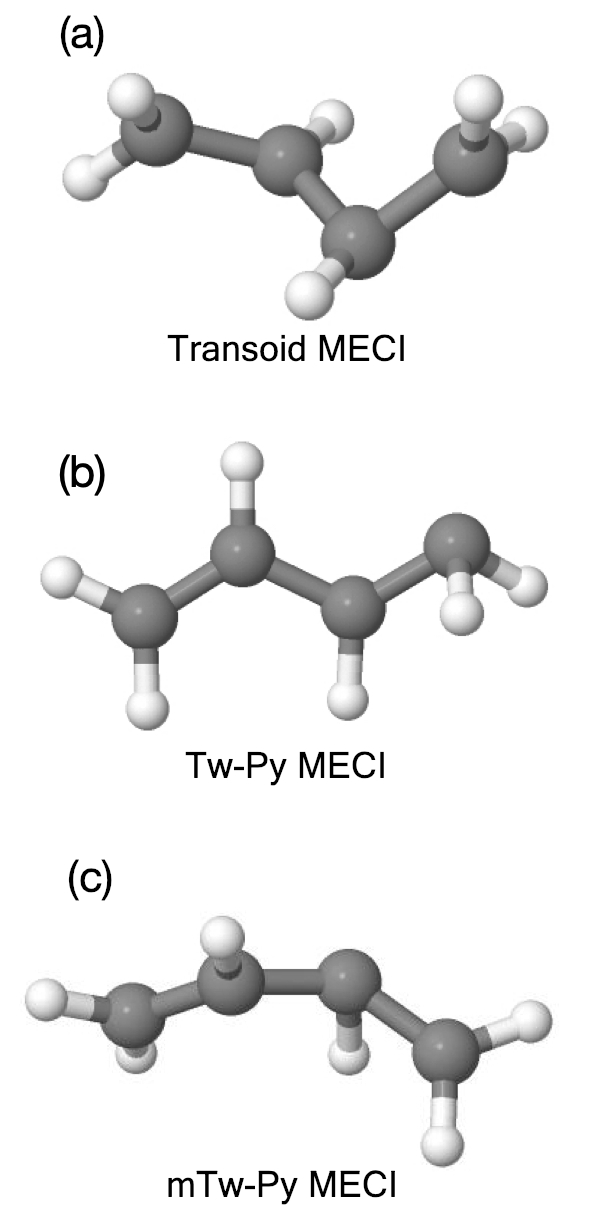}
    \caption{Minimum energy conical intersection (MECI) geometries, (a) Transoid (b) Tw-Py and (c) medial Tw-Py (mTw-Py).}
    \label{fig:meci_geom}
\end{figure}

Figure \ref{fig:adpop} shows the time-dependent adiabatic state populations, which are the same as those shown in Figure 2(a) in Ref. \cite{MacDonell_JPhysChemA_2019}. Each trajectory basis function was initialized on the adiabatic surface corresponding to the optically bright $\pi\pi^*$ state. At this level of \textit{ab initio} theory (and sampling of the ground-state vibrational distribution), this is predominantly the $S_2$ state, although $\sim 5\%$ of trajectories were started on $S_1$. As the Figure shows, the total excited-state population decays with a time constant of roughly $\tau \approx 210$ fs.

Previous studies have interpreted the ultrafast relaxation to the ground electronic state primarily in terms of nuclear structural motifs. Specifically, ethylenic conical intersections involving large-amplitude torsion about a C-C bond (generally a terminal C-C bond), followed by pyramidalization (termed twisted-pyramidalized, Tw-Py), characteristic of the establishment of polarization across the C-C bond. Minor channels have previously been identified (and are observed here again) involving pyramidalization at a medial, rather than terminal carbon atom. The other primary motif, previously termed ``transoid",\cite{Levine_JPhysChemA_2009} involves torsion about the medial C-C bond, heralding excited-state initiation of cis-trans isomerization. In contrast to the Tw-Py nuclear configurations, these structures do not display significant zwitterionic character, rather, multi-radical electronic structures are realized across the C-C-C-C backbone. The nuclear structures of the Tw-Py (both terminal and medial, the latter denoted mTw-Py) and transoid optimized MECIs are shown in Figure \ref{fig:meci_geom}. The cartesian coordinates of all reported nuclear structures are given as supplementary information in Tables S3 - S7. 

The relative importance of these structures formed the basis of much analysis in previous studies,\cite{Levine_JPhysChemA_2009,Glover_JChemPhys_2018,MacDonell_JPhysChemA_2019,MacDonell_JChemPhys_2020} including the ability to affect the branching between the Tw-Py and transoid motifs using chemical substitution\cite{MacDonell_JChemPhys_2020}. However, as discussed in that previous work, characterizing nuclear structures as uniquely ``Tw-Py'' or ``transoid'' is often difficult, as the large-amplitude nuclear motion often involves displacements along both of these directions. Therefore, the present analysis will focus instead on the \textit{electronic} structures associated with the CI-mediated electronic relaxation pathways.

\begin{figure}[hbt!]
\centering
    \includegraphics[width=.6\textwidth]{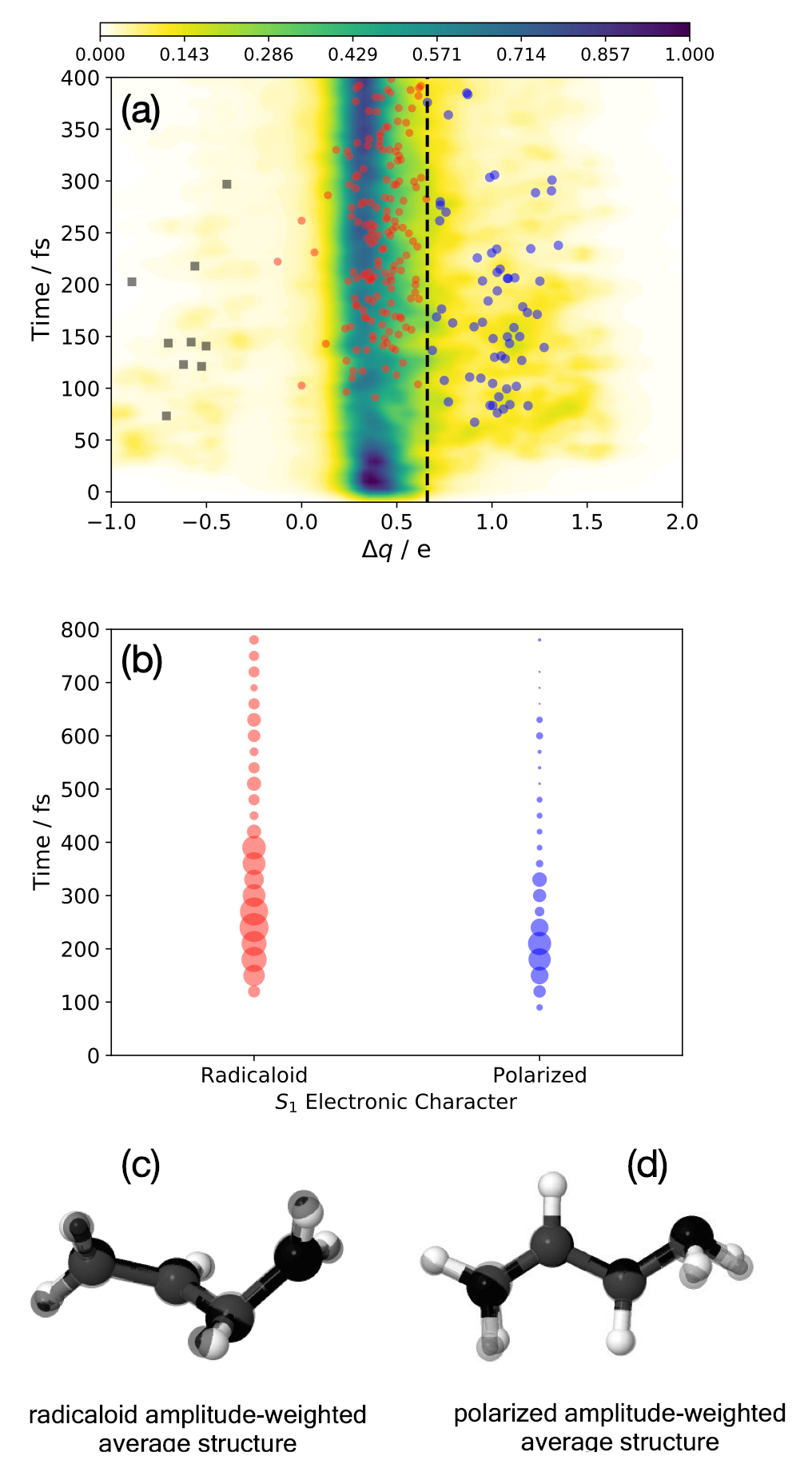}
    \caption{The evolution of the electronic character of the wave packet. The amplitude-weighted polarization across the terminal C-C bond is shown in (a), $\Delta q \approx 0\:(1)$ corresponds to radicaloid (polarized) structures. Spawn points are denoted by red (radicaloid) and blue circles (Tw-Py/polarized) and black squares (mTw-Py/polarized). The $S_1 \rightarrow S_0$ spawn points are binned by electronic character in (b), where the circle radius is proportional to the population transferred. Amplitude-weighted average nuclear structures (in black) are shown for the red (c) and blue points (d). These structures are superimposed on the transoid and Tw-Py MECI shown in Figure 1.}
    \label{fig:elstruct_evolution2}
\end{figure}

Previous work has examined multiple techniques for evaluating atomic partial charges at nuclear configurations near conical intersections and employing the multi-reference electronic structures that characterize these regions of coordinate space\cite{MacDonell_ChemPhys_2018}. The iterative Hirshfeld approach was found to yield intuitive results that were also only weakly dependent on the level of theory employed\cite{MacDonell_JChemPhys_2020}. To quantify the degree of polarization between a terminal and medial carbon atom, these charges were computed for each sampled trajectory and at each time step, and the difference between these two sites were determined. The larger of the two charge differences (i.e. the largest magnitude difference of the quantity $q_{\text{medial}}-q_{\text{terminal}}$, where the sign of the difference is maintained) will subsequently be denoted ``$\Delta q$". Structures evincing highly polarized characters gives rise to values of $\Delta q \approx 1$, whereas those structures with negligible charge separation yield $\Delta q \approx 0$. Figure \ref{fig:elstruct_evolution2}(a) plots these values for each trajectory basis function, at each time, weighted by the corresponding amplitude. The bifurcation of the wave packet into `radicaloid' and `polarized' components is clearly shown by the Figure \ref{fig:elstruct_evolution2}(a), with (the predominant) branch centered about $\Delta q = 0.3$ corresponding to the former, and the branch centered about $\Delta q = 1$ corresponding to the latter. The spawn events are superimposed as circles on the charge plot in Fig. \ref{fig:elstruct_evolution2}(a). Structures we denote as radicaloid or polarized spawns are colored red and blue, respectively. Interestingly, this plot also illustrates the presence of the third nuclear motif, involving pyramidalization at a medial (rather than terminal carbon atom). The corresponding negative charge at the medial carbon gives rise to a negative $\Delta q$ and are identified with the black squares in Fig. \ref{fig:elstruct_evolution2}(a). 

A further analysis of the $S_1 \rightarrow S_0$ spawn geometries is shown in Figure \ref{fig:elstruct_evolution2}(b), in which the electronic characters at these points form the basis for distinguishing the relevant region of the CI seam. Specifically: the spawn events are ``binned" according to the electronic character of the $S_1$ state as well as in 25 fs time intervals, while the radius of the circles in Fig. \ref{fig:elstruct_evolution2}(b) are proportional to the population spawned in each interval. We see that in the present dynamics simulation, electronic relaxation to the ground-state occurs primarily via radicaloid electronic structures, while the polarized structures are a minor channel.

While this analysis, which focuses on the electronic character of the wave packet, is highly illustrative (and relatively unambiguous), some consideration of the nuclear structures associated with the electronic decay is warranted. Figs.~\ref{fig:elstruct_evolution2}(c) and (d) plot the ``average'' spawn geometries for the radicaloid (c) and polarized (d) spawn geometries. These structures are the amplitude-weighted average Cartesian spawn geometries for radicaloid and polarized ``bins'', and which were determined only after the structures were placed into maximum coincidence\cite{Kabsch_ActaCryst_1976}. The small number of mTw-Py spawn events, as well as the more extended seam region sampled by these points, precluded this simple analysis for the spawn events denoted by the black squares in Fig.~\ref{fig:elstruct_evolution2}(a). The similarity of the structures in Figs.~\ref{fig:elstruct_evolution2}(c) and (d) (in all black) to the MECIs in Figs. \ref{fig:meci_geom}(b) and (c), (which are superimposed with normal coloring), is striking. We see that the polarized structures correspond to the previously identified Tw-Py structures, while the radicaloid structures are characterized by large amplitude dihedral motion about the C-C-C-C backbone as observed for the transoid MECI.

\subsection{Static X-ray Absorption Spectra}

Before turning to the simulation of the time-resolved spectrum, we present first the static XAS at key nuclear configurations from the relevant electronic states. Figure \ref{fig:fc_xas} shows the static XAS at the (a) Franck-Condon (FC) point, (b) Tw-Py $S_1 - S_0$ MECI, and (c) transoid $S_1 - S_0$ MECI.

Fig.~\ref{fig:fc_xas}(a) presents three spectra, each computed at the ground-state minimum energy nuclear configuration: the XAS from the ground-state ($S_0$), optically dark electronic state (denoted here as the $S_1$ ($\pi^{*2}$)), and bright $\pi\pi^*$ ($S_2$). For the $S_2$ XAS at the FC point, we observe four main peaks, and only one band near 283.5 eV about $\sim 1 eV$ to the red of the ground-state absorption, has appreciable intensity.  An examination of the NTOs, shown in Table \ref{tab:fc_nto}, for the underlying transitions revealed that the first two peaks correspond (A$''$ and B$''$) to transitions into the unoccupied $\pi$-orbital, with a splitting of $\sim$0.4 eV between the pairs of $1s$ orbitals localised on the terminal and medial carbon atoms. The other two peaks (including the highest intensity peak D$''$) are dominated by transitions into the Rydberg manifold. The $S_1$ state has $\pi^{*2}$ character and is optically dark. The two lowest intensity peaks (A$'$, B$'$) for this initial state are similar in character to the $S_2$ NTOs for the same lowest energy transitions, but are the $++$-combination of the $1s$ orbitals. The corresponding transitions of the $+-$-combination are found at higher energy and characterize peaks C$'$ and D$'$, with this pair forming the most intense transitions in the spectrum. The $S_0$ XAS evinces five high intensity peaks, labeled A, B, C, D, and E. Peaks A and B, the two highest intensity transitions in the spectrum, correspond to excitation from different C1s orbitals into the LUMO $\pi^*$, with the latter peak displaying mixed valence-Rydberg character. The remaining high intensity peaks involve transitions into orbitals of Rydberg character. In summary, the majority of the strongest intensity pre-edge transitions for all three states involve core-transitions into the $\pi$ manifold; the empty $\pi^*$ for the ground-state, and into a hole in a $\pi$ orbital for the electronically excited states. The highest intensity peak in the $S_2$ XAS, however, corresponds to a transition between C1s and 3d Rydberg orbitals.

Anticipating their relevance in the assignment of the spectral features in the TRXAS, we also show simulated XAS for the $S_0$ and $S_1$ states at the Tw-Py and transoid $S_1 - S_0$ MECIs in Figure \ref{fig:fc_xas}(b) and (c), respectively. Since the adiabatic labels $S_0$ and $S_1$ are arbitrary at a point of degeneracy, we instead denote the states as polarized (pol.) or radicaloid (rad.) in Fig. \ref{fig:fc_xas}(b). Neither electronic state displays significant polarization at the transoid MECI, so the states are labeled ``rad.~1'' and ``rad.~2'', where the former corresponds to the $S_1$ electronic character upon approach to the MECI from the FC region. The NTOs for these spectra are presented as supplementary material. Figure \ref{fig:fc_xas}(b) shows a low intensity peak A, at 283.0 eV, significantly red-shifted from a higher intensity doublet centered (peaks B and C) around $\sim 284.5$eV. As Table S2 shows, peak A corresponds to a transition from the pyramidalized C $1s$ orbital to the hole in $\pi$-type orbital. The large partial negative charge localised around the pyramidalised C atom 
induces a large chemical shift in a manner analogous to that observed previously in a simulated TRXAS of ethylene\cite{Neville_ADC_PhysRevLett_2018}. Peaks B and C arise from the transitions from the other C $1s$ orbitals on the backbone into $\pi$-type orbitals. The spectrum arising from the radicaloid electronic state displays little overlap with that of the polarized state, with peaks A-D again arising from transitions into $\pi$-type orbitals. Conversely, the spectra for the rad.~1 and rad.~2 states, computed at the transoid MECI and shown in Fig. \ref{fig:fc_xas}(c), display a high degree of overlap. However, the main spectral features are red-shifted ($\sim$ 283.5 eV vs. $\sim$ 284.5 eV) relative to the high intensity peaks B and C for the polarized electronic state at the Tw-Py, suggesting the two structural motifs are potentially resolvable via their distinct XAS.

\subsection{Simulation of the Time-Resolved X-ray Absorption Spectrum}

In the absence of experimental results, we have chosen to present the isotropic spectrum, the polarization dependent spectra are given as supplementary material in Figs.~S2 and S3. The perpendicular pump-probe relative polarization configuration is predicted to yield signal an order of magnitude higher than the parallel polarization and is nearly identical to the isotropic simulation. Figure \ref{fig:trxas_iso} shows the full time-resolved spectrum, which employs a global energy shift, for all initial states, of 3.5 eV. This value was chosen so as to align the main peaks in ground-state XAS, see Fig.~S1. 

\begin{figure}[hbt!]
    \centering
    \includegraphics[width=.5\textwidth]{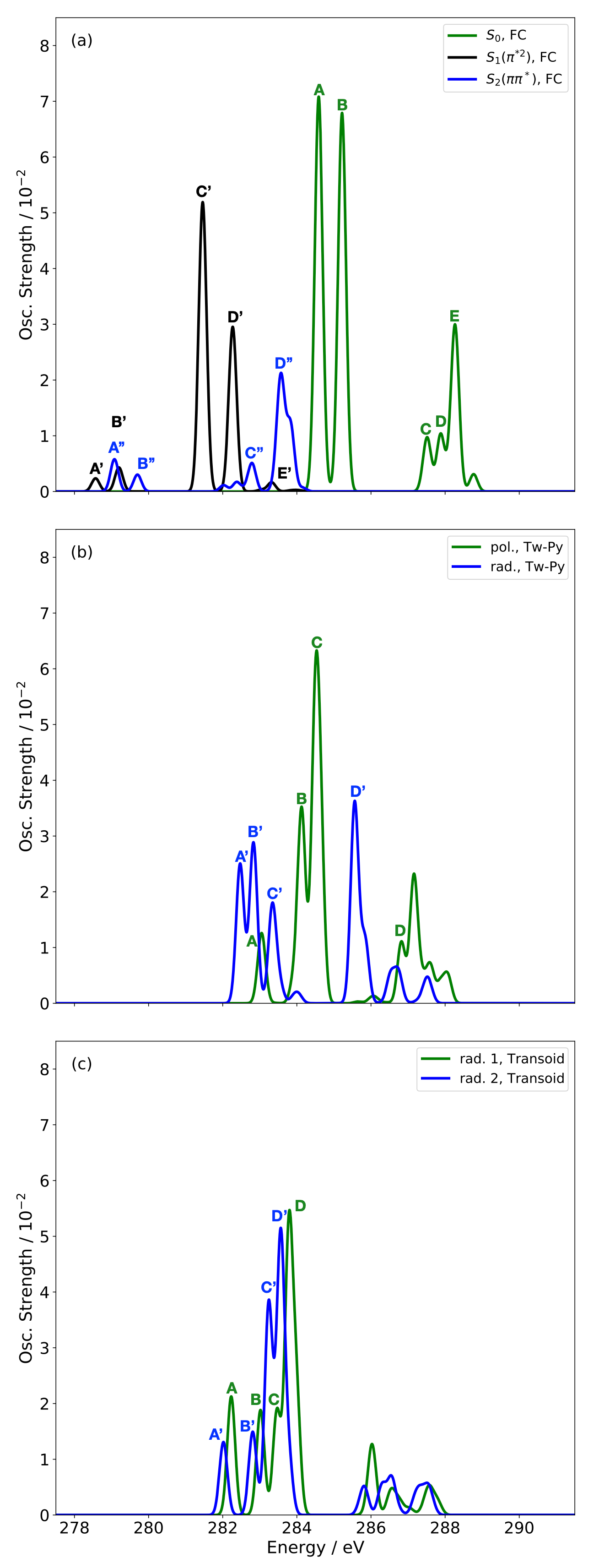}
    \caption{Simulated XAS at CVS-DFT/MRCI/aug-cc-pVDZ level of theory for $S_0$ , $S_1$ and $S_2$ states at the FC (a) and key MECI geometries (Tw-Py (b) and Transoid (c)) of 1,3-BD.}
    \label{fig:fc_xas}
\end{figure}

\begin{figure}[hbt!]
    \centering
    \includegraphics[width=1.\textwidth]{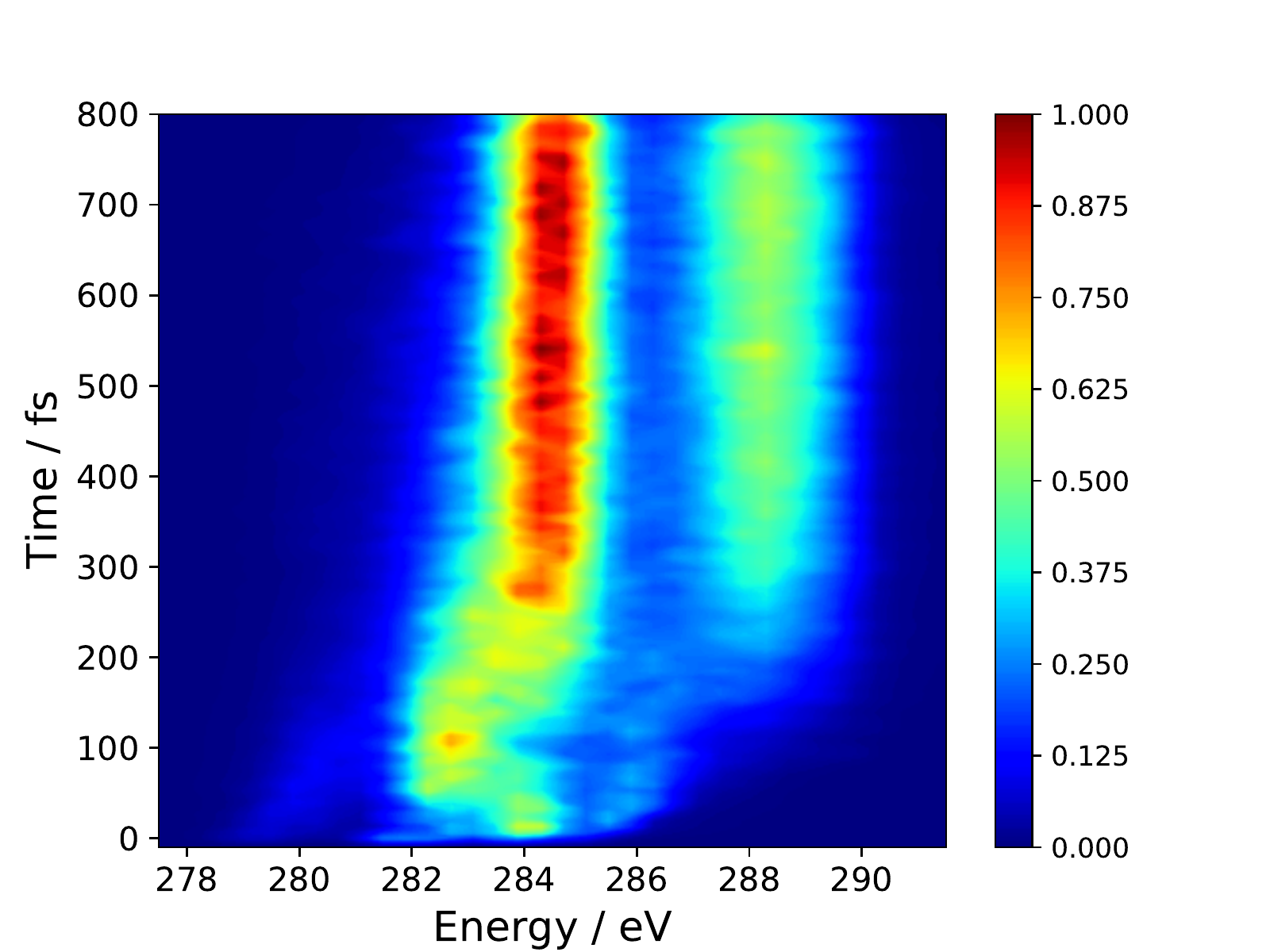}
    \caption{Simulated TRXAS spectrum employing the results of AIMS dynamics simulations and CVS-DFT/MRCI X-ray absorption cross-sections. An isotropic axis distribution was assumed.}
    \label{fig:trxas_iso}
\end{figure}

The spectral features present at time, $t=0$, correspond to the peaks present in the static $S_2$ XAS at the FC point (Peaks A$''$, B$''$, C$''$, D$''$). In particular, the high intensity peak at $\sim 284.0$ corresponds to the high intensity peak labeled D$''$ in Fig. \ref{fig:fc_xas}(a) for the $S_2$ state. This transition is assigned to an excitation of the 1s orbitals on the terminal C-atoms into a 3d Rydberg orbital. 

\begin{figure}[hbt!]
    \centering
    \includegraphics[width=.4\textwidth]{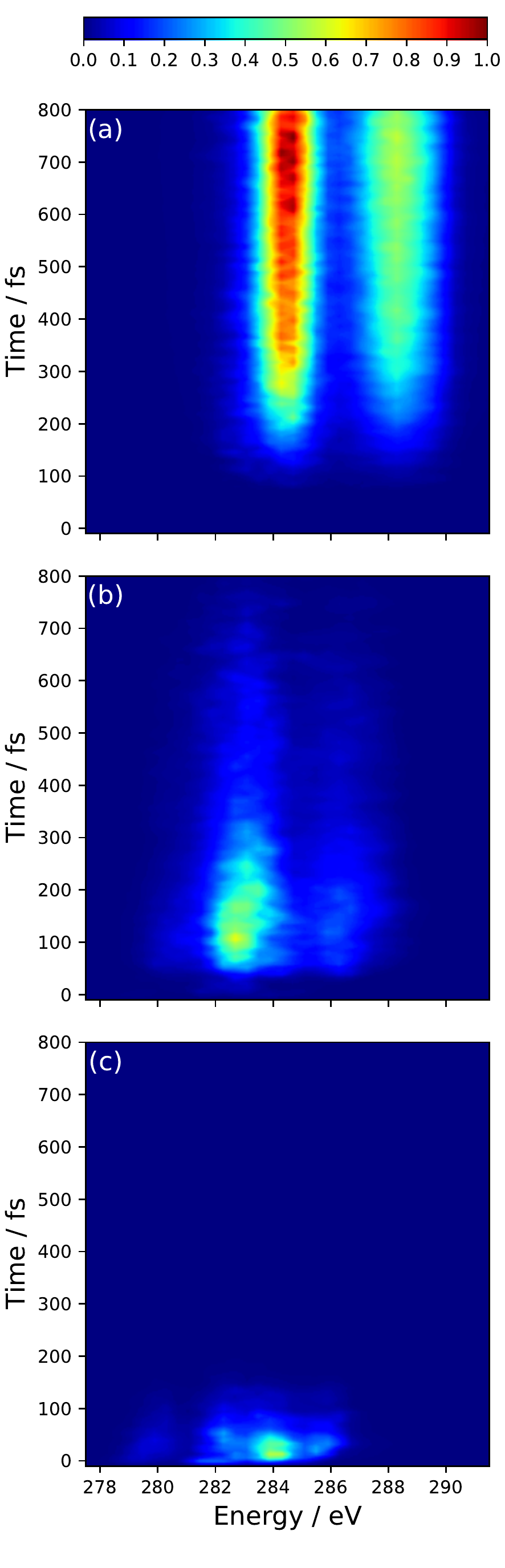}
    \caption{Simulated TRXAS at CVS-DFT/MRCI/aug-cc-pVDZ level of theory, decomposed into contributions from the (a) $S_2$, (b) $S_1$ and (c) $S_0$ adiabatic states}
    \label{fig:trxas_s0s1s2}
\end{figure}

This spectral feature sweeps to higher energy with a period of $\sim$50 fs and tracks the electronic relaxation from the $S_2$ to the $S_1$ state. Support for this assignment is evinced by the decomposition of the TRXAS into contributions from the adiabatic states shown in Fig.~\ref{fig:trxas_s0s1s2}. The spectral features in the static $S_1$ XAS shown in Fig.~\ref{fig:fc_xas} map onto the shorter time features in Fig.~\ref{fig:trxas_s0s1s2}(b), assigned to $S_1$ adiabatic state contribution.  This change in adiabatic state would appear to also correspond to a change of the wave packet to the ``dark'' $\pi^{*2}$ electronic character. At short times, prior to large amplitude motion about the C-C-C-C backbone, Fig.\ref{fig:fc_xas}(a) shows that the XAS transitions from the $\pi^{*2}$ state are red-shifted to those of the initially prepared $\pi\pi^*$ state by $1.5-2$ eV (i.e. peaks C$'$ and D$'$). Furthermore, at longer times, the analysis of the dynamics results show that the transoid structures are the primary relaxation nuclear motif. Fig.\ref{fig:fc_xas}(c) shows that the radicaloid electronic structures lead to XAS transitions at energies less than 283.5 eV, while the high intensity absorption lines in the XAS for polarized electronic structures occur at energies greater than 283.5 eV (i.e.~peaks B and C in Fig.\ref{fig:fc_xas}(b)). 

The spectral features that characterize the TRXAS at longer times (> 300fs) correspond to the return of the wave packet to the vibrationally hot ground-state, as shown in Fig.~\ref{fig:trxas_s0s1s2}(c). These spectral features, though dramatically broadened, align well with peaks A and B for the band around 284.5 eV and peaks C-E for the band above 287.5 eV.

\begin{figure}[hbt!]
    \centering
    \includegraphics[width=.9\textwidth]{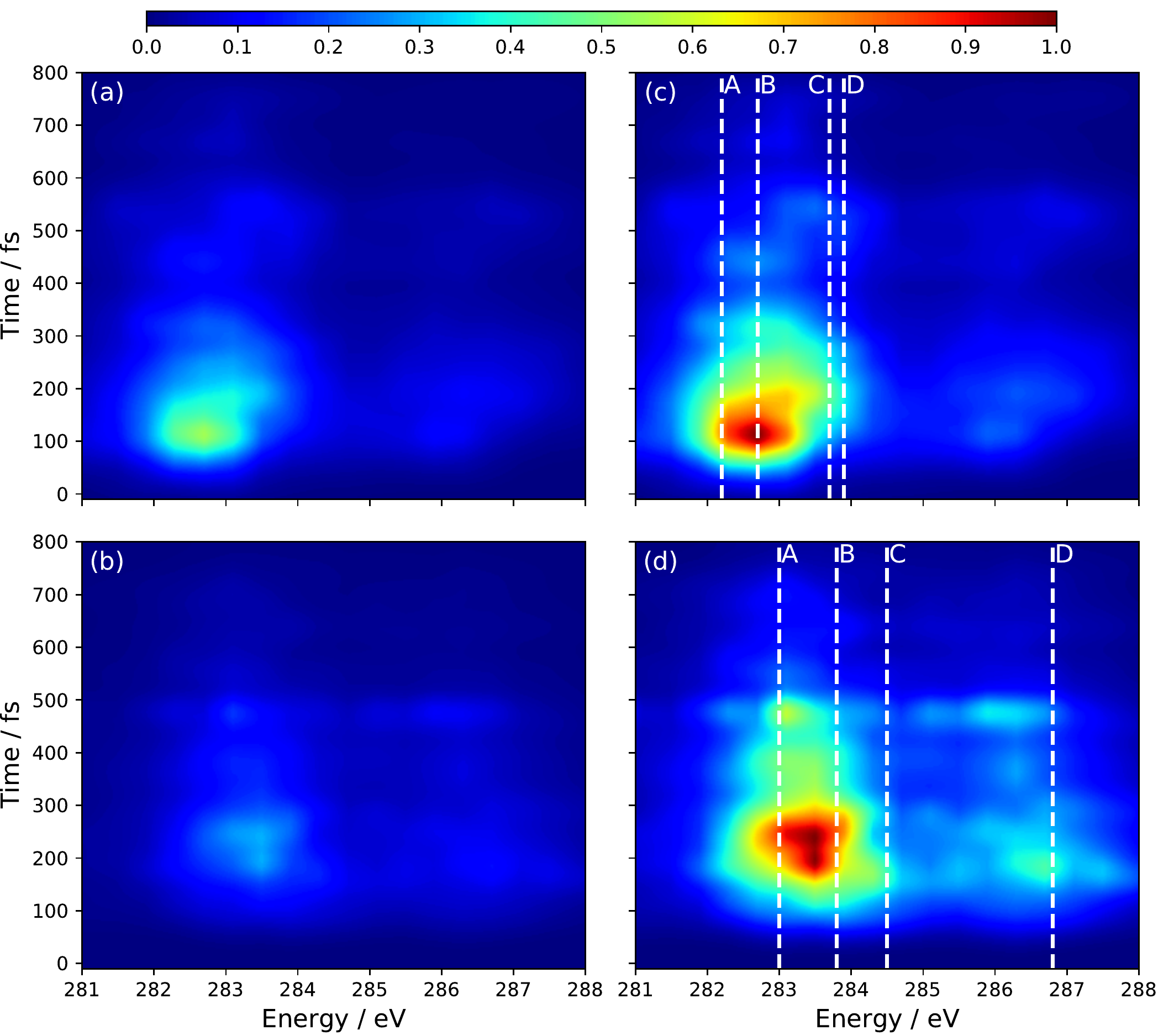}
    \caption{Simulated $S_1$ TRXAS at CVS-DFT/MRCI/aug-cc-pVDZ level of theory, decomposed into radicaloid (a) and polarized (b) contributions. (c) and (d) correspond to the intensity-renormalized plot of (a) and (b), respectively.}
    \label{fig:s1_trxas_s1_radpol}
\end{figure}

The remaining question is to what extent the $S_1$ component to the TRXAS spectrum to the spectrum can be decomposed into contributions from the two electronic configurations associated with the polarized and radicaloid decay pathways. Using the bifurcation of the wave packet evinced by Fig.~\ref{fig:elstruct_evolution2}, we decompose the $S_1$ adiabatic state contribution to the TRXAS (Fig.~\ref{fig:trxas_s0s1s2}(b)) into contributions from those trajectory basis functions with $\Delta q < 0.55$ (radicaloid) and $\Delta q > 0.55$ (polarized). Figure \ref{fig:s1_trxas_s1_radpol} shows the decomposed spectra. Panels (a) and (b) show the two contributions where the normalization of the intensity is chosen to be consistent with Fig.~\ref{fig:trxas_s0s1s2} so that the sum of the contributions yield that spectrum. Panels (c) and (d) re-normalize the intensity of each individual contribution to aid in the visualization of the predicted spectral signal from each electronic character.  

Consistent with our analysis of the spawned geometries, as well as the bifurcation of the wave packet 
into components with radicaloid and polarized characters, a comparison of the relative intensities in Figs.~\ref{fig:s1_trxas_s1_radpol}(a) and (b) shows that the spectrum is dominated by the component of the wave packet with radicaloid character. 
Indeed, the (broad) bands around 283.5 eV and 286.5 eV in Figs.~\ref{fig:s1_trxas_s1_radpol}(a) and (c) correspond closely to the peaks in the static XAS in Fig.~\ref{fig:fc_xas}(c) computed at the transoid MECI. The excitation energies of the main peaks in the in the static XAS in Fig.~\ref{fig:fc_xas}(c) are shown as dashed lines (where the line labels correspond to the peak labels in Fig.~\ref{fig:fc_xas}) to facilitate the comparison.  

While the signal assigned to polarized structures overlaps strongly with the radicaloid signal, the dominant main bands in the component spectra in \ref{fig:s1_trxas_s1_radpol}(b) and (d) (284.5 eV and 287.5 eV) are both blue-shifted by $\sim 1$eV relative to the radicaloid peaks. Furthermore, these bands corresponds closely to the peaks in the static spectrum for the polarized electronic state at the Tw-Py structure in Fig.\ref{fig:fc_xas}(b). The dominant structural feature in the polarized-component TRXAS lies between peaks A and B in the static spectrum, due to significant spectral broadening arising from the significant vibrational energy on the $S_1$ surface. Peak D in the static spectrum, though, is clearly discernable in Fig.~\ref{fig:s1_trxas_s1_radpol}(d). That said, the main issue in resolving these pathways is that the polarized component of the TRXAS is of significantly lower intensity since the transoid CI relaxation channel is dominant in the present dynamics simulation. 

However, previous work\cite{MacDonell_ChemPhys_2018, MacDonell_JPhysChemA_2019} has shown that the branching between these paths can be influenced via chemical substitution at specific sites on the C-backbone. This can be achieved either by methyl-group or functional group substitution. The former, when employed at the terminal C-atoms\cite{MacDonell_JChemPhys_2020}, has been shown to further reduce the prevalence of the polarized structures. Conversely, substitution of the $\pi$-donor nitrile group has been predicted, via simulation,\cite{MacDonell_JPhysChemA_2019} to dramatically lower the energy of polarized structures when placed at specific points on the C-C-C-C backbone. The effect of chemical substitution, and thus the modified non-adiabatic dynamics, on the simulated TRXAS will be the basis for a future work. 

\section{Conclusion}

On the basis of high-level nuclear dynamics simulations and quantum chemical computations, we here demonstrate the ability of time-resolved X-ray absorption spectroscopy to resolve the electronic structures corresponding to the different conical intersections that are observed in the excited-state dynamics of trans 1,3-BD. When a wave packet is initialized on the lowest-energy bright electronic state in the FC region, the resulting non-adiabatic dynamics ultimately result in a return to the ground-state via two excited-state electronic structures: a tetraradical species associated with nuclear structures along the trans-cis isomerization coordinate, and a highly polarized transient structure analogous to the Tw-Py MECI of ethylene. Employing an \textit{ab initio} multiple spawning dynamics simulation, and the component trajectory basis functions, we simulated the TRXAS using the recently developed CVS-DFT/MRCI method, which is capable of describing the electronic structures that result from core-excitation of highly multi-reference initial valence states. The TRXAS was then decomposed into contributions arising from not only each adiabatic state, but also from the radicaloid and polarized electronic components of the total excited-state wave packet. We found that while the initially prepared $S_2$ is unambiguously of bright $\pi\pi^*$ character, the wave packet on $S_1$ was, unsurprisingly a mixture of nuclear and electronic structures.  Decomposition of the TRXAS into contributions in terms of the radicaloid and polarized structures revealed that the key spectral features could be resolved and that the present simulation strongly favors the transoid structures for electronic relaxation to the ground-state. 

We close by noting that although the dynamics of 1,3-BD (simulated using the particular potential energy surfaces of the present study) strongly favors one of the two pathways, this result can be influenced via selective chemical substitution. Previous work has shown that selective stabilization (de-stabilization) of the polarized electronic structures associated with the Tw-Py can inhibit (promote) the cis-trans isomerization pathway. Tuning this branching between the different dynamical pathways, and simulating the associated time-resolved core spectroscopies to assess their utility in discerning the different electronic and nuclear structures, will serve as the basis for future work.

\section{Acknowlegements}
The authors thank A.~Stolow for helpful comments during the preparation of this manuscript. MSS acknowledges the financial support of Natural Sciences and Engineering Research Council (NSERC) through the Discovery Grant program.

\clearpage

\begin{table}[htb!]
  \caption{FC NTOs of the ground-state and excited-state XAS.}
  \begin{center}
    \includegraphics[width=.9\textwidth]{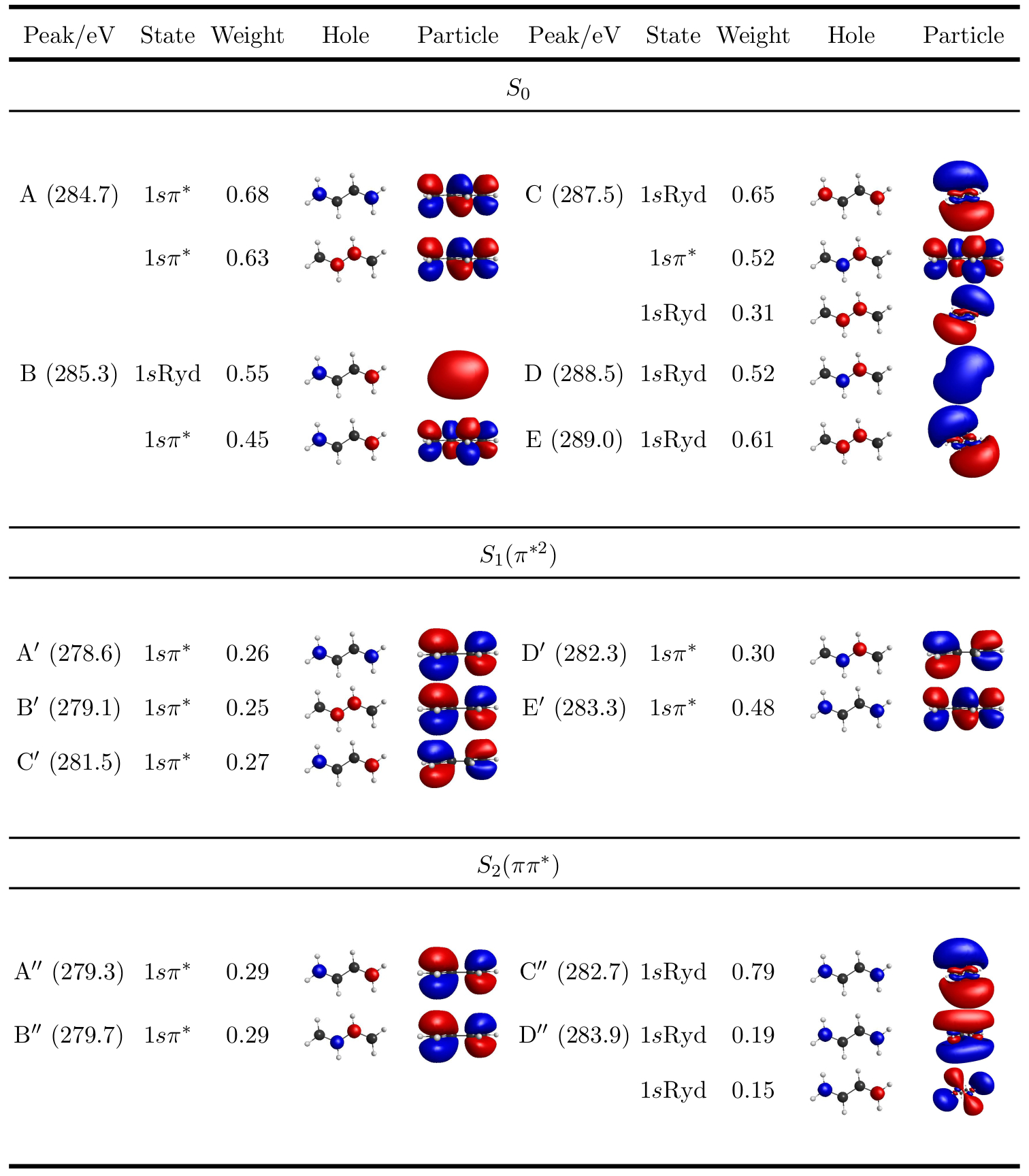}
  \end{center}
  \label{tab:fc_nto}
\end{table}

\newpage
\clearpage
\bibliography{references}

\begin{thebibliography}{46}
\expandafter\ifx\csname natexlab\endcsname\relax\def\natexlab#1{#1}\fi
\expandafter\ifx\csname bibnamefont\endcsname\relax
  \def\bibnamefont#1{#1}\fi
\expandafter\ifx\csname bibfnamefont\endcsname\relax
  \def\bibfnamefont#1{#1}\fi
\expandafter\ifx\csname citenamefont\endcsname\relax
  \def\citenamefont#1{#1}\fi
\expandafter\ifx\csname url\endcsname\relax
  \def\url#1{\texttt{#1}}\fi
\expandafter\ifx\csname urlprefix\endcsname\relax\def\urlprefix{URL }\fi
\providecommand{\bibinfo}[2]{#2}
\providecommand{\eprint}[2][]{\url{#2}}

\bibitem[{\citenamefont{Blanchet et~al.}(1999)\citenamefont{Blanchet, Zgierski,
  Seideman, and Stolow}}]{Blanchet_Nature_1999}
\bibinfo{author}{\bibfnamefont{V.}~\bibnamefont{Blanchet}},
  \bibinfo{author}{\bibfnamefont{M.~Z.} \bibnamefont{Zgierski}},
  \bibinfo{author}{\bibfnamefont{T.}~\bibnamefont{Seideman}}, \bibnamefont{and}
  \bibinfo{author}{\bibfnamefont{A.}~\bibnamefont{Stolow}},
  \bibinfo{journal}{Nature} \textbf{\bibinfo{volume}{401}}, \bibinfo{pages}{52}
  (\bibinfo{year}{1999}), ISSN \bibinfo{issn}{1476-4687},
  \urlprefix\url{https://doi.org/10.1038/43410}.

\bibitem[{\citenamefont{Stolow}(2003)}]{Stolow_AnnuRevPhysChem_2003}
\bibinfo{author}{\bibfnamefont{A.}~\bibnamefont{Stolow}},
  \bibinfo{journal}{Annu. Rev. Phys. Chem.} \textbf{\bibinfo{volume}{54}},
  \bibinfo{pages}{89} (\bibinfo{year}{2003}), \bibinfo{note}{pMID: 12524428},
  \eprint{https://doi.org/10.1146/annurev.physchem.54.011002.103809},
  \urlprefix\url{https://doi.org/10.1146/annurev.physchem.54.011002.103809}.

\bibitem[{\citenamefont{Stolow et~al.}(2004)\citenamefont{Stolow, Bragg, and
  Neumark}}]{Stolow_ChemRev_2004}
\bibinfo{author}{\bibfnamefont{A.}~\bibnamefont{Stolow}},
  \bibinfo{author}{\bibfnamefont{A.~E.} \bibnamefont{Bragg}}, \bibnamefont{and}
  \bibinfo{author}{\bibfnamefont{D.~M.} \bibnamefont{Neumark}},
  \bibinfo{journal}{Chem. Rev.} \textbf{\bibinfo{volume}{104}},
  \bibinfo{pages}{1719} (\bibinfo{year}{2004}), \bibinfo{note}{pMID: 15080710},
  \eprint{https://doi.org/10.1021/cr020683w},
  \urlprefix\url{https://doi.org/10.1021/cr020683w}.

\bibitem[{\citenamefont{Hofmann and
  de~Vivie-Riedle}(2001)}]{Hofmann_ChemPhysLett_2001}
\bibinfo{author}{\bibfnamefont{A.}~\bibnamefont{Hofmann}} \bibnamefont{and}
  \bibinfo{author}{\bibfnamefont{R.}~\bibnamefont{de~Vivie-Riedle}},
  \bibinfo{journal}{Chem. Phys. Lett.} \textbf{\bibinfo{volume}{346}},
  \bibinfo{pages}{299} (\bibinfo{year}{2001}), ISSN \bibinfo{issn}{0009-2614},
  \urlprefix\url{http://www.sciencedirect.com/science/article/pii/S0009261401009228}.

\bibitem[{\citenamefont{Schuurman and
  Stolow}(2018)}]{Schuurman_AnnuRevPhysChem_2018}
\bibinfo{author}{\bibfnamefont{M.~S.} \bibnamefont{Schuurman}}
  \bibnamefont{and} \bibinfo{author}{\bibfnamefont{A.}~\bibnamefont{Stolow}},
  \bibinfo{journal}{Annu. Rev. Phys. Chem.} \textbf{\bibinfo{volume}{69}},
  \bibinfo{pages}{427} (\bibinfo{year}{2018}), \bibinfo{note}{pMID: 29490199},
  \eprint{https://doi.org/10.1146/annurev-physchem-052516-050721},
  \urlprefix\url{https://doi.org/10.1146/annurev-physchem-052516-050721}.

\bibitem[{\citenamefont{Voll et~al.}(2007)\citenamefont{Voll, Kerscher,
  Geppert, and de~Vivie-Riedle}}]{Voll_JPhotochemPhotobiol_2007}
\bibinfo{author}{\bibfnamefont{J.}~\bibnamefont{Voll}},
  \bibinfo{author}{\bibfnamefont{T.}~\bibnamefont{Kerscher}},
  \bibinfo{author}{\bibfnamefont{D.}~\bibnamefont{Geppert}}, \bibnamefont{and}
  \bibinfo{author}{\bibfnamefont{R.}~\bibnamefont{de~Vivie-Riedle}},
  \bibinfo{journal}{J. Photochem. Photobiol. A} \textbf{\bibinfo{volume}{190}},
  \bibinfo{pages}{352} (\bibinfo{year}{2007}), ISSN \bibinfo{issn}{1010-6030},
  \bibinfo{note}{theoretical Aspects of Photoinduced Processes in Complex
  Systems},
  \urlprefix\url{http://www.sciencedirect.com/science/article/pii/S1010603007000846}.

\bibitem[{\citenamefont{Geneaux et~al.}(2019)\citenamefont{Geneaux, Marroux,
  Guggenmos, Neumark, and Leone}}]{Geneaux_2019}
\bibinfo{author}{\bibfnamefont{R.}~\bibnamefont{Geneaux}},
  \bibinfo{author}{\bibfnamefont{H.~J.~B.} \bibnamefont{Marroux}},
  \bibinfo{author}{\bibfnamefont{A.}~\bibnamefont{Guggenmos}},
  \bibinfo{author}{\bibfnamefont{D.~M.} \bibnamefont{Neumark}},
  \bibnamefont{and} \bibinfo{author}{\bibfnamefont{S.~R.} \bibnamefont{Leone}},
  \bibinfo{journal}{Philos.~Trans.~Royal Soc.~A}
  \textbf{\bibinfo{volume}{377}}, \bibinfo{pages}{20170463}
  (\bibinfo{year}{2019}),
  \eprint{https://royalsocietypublishing.org/doi/pdf/10.1098/rsta.2017.0463},
  \urlprefix\url{https://royalsocietypublishing.org/doi/abs/10.1098/rsta.2017.0463}.

\bibitem[{\citenamefont{Boguslavskiy et~al.}(2018)\citenamefont{Boguslavskiy,
  Schalk, Gador, Glover, Mori, Schultz, Schuurman, Mart{\'i}nez, and
  Stolow}}]{Boguslavskiy_JChemPhys_2018}
\bibinfo{author}{\bibfnamefont{A.~E.} \bibnamefont{Boguslavskiy}},
  \bibinfo{author}{\bibfnamefont{O.}~\bibnamefont{Schalk}},
  \bibinfo{author}{\bibfnamefont{N.}~\bibnamefont{Gador}},
  \bibinfo{author}{\bibfnamefont{W.~J.} \bibnamefont{Glover}},
  \bibinfo{author}{\bibfnamefont{T.}~\bibnamefont{Mori}},
  \bibinfo{author}{\bibfnamefont{T.}~\bibnamefont{Schultz}},
  \bibinfo{author}{\bibfnamefont{M.~S.} \bibnamefont{Schuurman}},
  \bibinfo{author}{\bibfnamefont{T.~J.} \bibnamefont{Mart{\'i}nez}},
  \bibnamefont{and} \bibinfo{author}{\bibfnamefont{A.}~\bibnamefont{Stolow}},
  \bibinfo{journal}{J. Chem. Phys.} \textbf{\bibinfo{volume}{148}},
  \bibinfo{pages}{164302} (\bibinfo{year}{2018}),
  \eprint{https://doi.org/10.1063/1.5016452},
  \urlprefix\url{https://doi.org/10.1063/1.5016452}.

\bibitem[{\citenamefont{Hudock et~al.}(2007)\citenamefont{Hudock, Levine,
  Thompson, Satzger, Townsend, Gador, Ullrich, Stolow, and
  Martínez}}]{Hudock_2007}
\bibinfo{author}{\bibfnamefont{H.~R.} \bibnamefont{Hudock}},
  \bibinfo{author}{\bibfnamefont{B.~G.} \bibnamefont{Levine}},
  \bibinfo{author}{\bibfnamefont{A.~L.} \bibnamefont{Thompson}},
  \bibinfo{author}{\bibfnamefont{H.}~\bibnamefont{Satzger}},
  \bibinfo{author}{\bibfnamefont{D.}~\bibnamefont{Townsend}},
  \bibinfo{author}{\bibfnamefont{N.}~\bibnamefont{Gador}},
  \bibinfo{author}{\bibfnamefont{S.}~\bibnamefont{Ullrich}},
  \bibinfo{author}{\bibfnamefont{A.}~\bibnamefont{Stolow}}, \bibnamefont{and}
  \bibinfo{author}{\bibfnamefont{T.~J.} \bibnamefont{Martínez}},
  \bibinfo{journal}{The Journal of Physical Chemistry A}
  \textbf{\bibinfo{volume}{111}}, \bibinfo{pages}{8500} (\bibinfo{year}{2007}),
  \bibinfo{note}{pMID: 17685594}, \eprint{https://doi.org/10.1021/jp0723665},
  \urlprefix\url{https://doi.org/10.1021/jp0723665}.

\bibitem[{\citenamefont{Glover et~al.}(2018)\citenamefont{Glover, Mori,
  Schuurman, Boguslavskiy, Schalk, Stolow, and
  Mart{\'i}nez}}]{Glover_JChemPhys_2018}
\bibinfo{author}{\bibfnamefont{W.~J.} \bibnamefont{Glover}},
  \bibinfo{author}{\bibfnamefont{T.}~\bibnamefont{Mori}},
  \bibinfo{author}{\bibfnamefont{M.~S.} \bibnamefont{Schuurman}},
  \bibinfo{author}{\bibfnamefont{A.~E.} \bibnamefont{Boguslavskiy}},
  \bibinfo{author}{\bibfnamefont{O.}~\bibnamefont{Schalk}},
  \bibinfo{author}{\bibfnamefont{A.}~\bibnamefont{Stolow}}, \bibnamefont{and}
  \bibinfo{author}{\bibfnamefont{T.~J.} \bibnamefont{Mart{\'i}nez}},
  \bibinfo{journal}{J. Chem. Phys.} \textbf{\bibinfo{volume}{148}},
  \bibinfo{pages}{164303} (\bibinfo{year}{2018}),
  \eprint{https://doi.org/10.1063/1.5018130},
  \urlprefix\url{https://doi.org/10.1063/1.5018130}.

\bibitem[{\citenamefont{MacDonell and
  Schuurman}(2019)}]{MacDonell_JPhysChemA_2019}
\bibinfo{author}{\bibfnamefont{R.~J.} \bibnamefont{MacDonell}}
  \bibnamefont{and} \bibinfo{author}{\bibfnamefont{M.~S.}
  \bibnamefont{Schuurman}}, \bibinfo{journal}{J. Phys. Chem. A}
  \textbf{\bibinfo{volume}{123}}, \bibinfo{pages}{4693} (\bibinfo{year}{2019}),
  \bibinfo{note}{pMID: 31050897},
  \eprint{https://doi.org/10.1021/acs.jpca.9b02446},
  \urlprefix\url{https://doi.org/10.1021/acs.jpca.9b02446}.

\bibitem[{\citenamefont{Tsuru et~al.}(2021)\citenamefont{Tsuru, Vidal,
  P{\'a}pai, Krylov, M{\o}ller, and Coriani}}]{Tsuru_StructDyn_2021}
\bibinfo{author}{\bibfnamefont{S.}~\bibnamefont{Tsuru}},
  \bibinfo{author}{\bibfnamefont{M.~L.} \bibnamefont{Vidal}},
  \bibinfo{author}{\bibfnamefont{M.}~\bibnamefont{P{\'a}pai}},
  \bibinfo{author}{\bibfnamefont{A.~I.} \bibnamefont{Krylov}},
  \bibinfo{author}{\bibfnamefont{K.~B.} \bibnamefont{M{\o}ller}},
  \bibnamefont{and} \bibinfo{author}{\bibfnamefont{S.}~\bibnamefont{Coriani}},
  \bibinfo{journal}{Struct.~Dyn.} \textbf{\bibinfo{volume}{8}},
  \bibinfo{pages}{024101} (\bibinfo{year}{2021}),
  \eprint{https://doi.org/10.1063/4.0000070},
  \urlprefix\url{https://doi.org/10.1063/4.0000070}.

\bibitem[{\citenamefont{Golubev et~al.}(2021)\citenamefont{Golubev,
  Van{\'i}\v{c}ek, and Kuleff}}]{Golubev_PhysRevLett_2021}
\bibinfo{author}{\bibfnamefont{N.~V.} \bibnamefont{Golubev}},
  \bibinfo{author}{\bibfnamefont{J.}~\bibnamefont{Van{\'i}\v{c}ek}},
  \bibnamefont{and} \bibinfo{author}{\bibfnamefont{A.~I.}
  \bibnamefont{Kuleff}}, \bibinfo{journal}{Phys.~Rev.~Lett.}
  \textbf{\bibinfo{volume}{127}}, \bibinfo{pages}{123001}
  (\bibinfo{year}{2021}),
  \urlprefix\url{https://link.aps.org/doi/10.1103/PhysRevLett.127.123001}.

\bibitem[{\citenamefont{Northey et~al.}(2020)\citenamefont{Northey, Norell,
  Fouda, Besley, Odelius, and Penfold}}]{Northey_PhysChemChemPhys_2020}
\bibinfo{author}{\bibfnamefont{T.}~\bibnamefont{Northey}},
  \bibinfo{author}{\bibfnamefont{J.}~\bibnamefont{Norell}},
  \bibinfo{author}{\bibfnamefont{A.~E.~A.} \bibnamefont{Fouda}},
  \bibinfo{author}{\bibfnamefont{N.~A.} \bibnamefont{Besley}},
  \bibinfo{author}{\bibfnamefont{M.}~\bibnamefont{Odelius}}, \bibnamefont{and}
  \bibinfo{author}{\bibfnamefont{T.~J.} \bibnamefont{Penfold}},
  \bibinfo{journal}{Phys. Chem. Chem. Phys.} \textbf{\bibinfo{volume}{22}},
  \bibinfo{pages}{2667} (\bibinfo{year}{2020}),
  \urlprefix\url{http://dx.doi.org/10.1039/C9CP03019K}.

\bibitem[{\citenamefont{Neville
  et~al.}(2018{\natexlab{a}})\citenamefont{Neville, Chergui, Stolow, and
  Schuurman}}]{Neville_PhysRevLett_2018}
\bibinfo{author}{\bibfnamefont{S.~P.} \bibnamefont{Neville}},
  \bibinfo{author}{\bibfnamefont{M.}~\bibnamefont{Chergui}},
  \bibinfo{author}{\bibfnamefont{A.}~\bibnamefont{Stolow}}, \bibnamefont{and}
  \bibinfo{author}{\bibfnamefont{M.~S.} \bibnamefont{Schuurman}},
  \bibinfo{journal}{Phys. Rev. Lett.} \textbf{\bibinfo{volume}{120}},
  \bibinfo{pages}{243001} (\bibinfo{year}{2018}{\natexlab{a}}),
  \urlprefix\url{https://link.aps.org/doi/10.1103/PhysRevLett.120.243001}.

\bibitem[{\citenamefont{Zinchenko et~al.}(2021)\citenamefont{Zinchenko,
  Ardana-Lamas, Seidu, Neville, van~der Veen, Lanfaloni, Schuurman, and
  W{\"o}rner}}]{Zinchenko_Science_2021}
\bibinfo{author}{\bibfnamefont{K.~S.} \bibnamefont{Zinchenko}},
  \bibinfo{author}{\bibfnamefont{F.}~\bibnamefont{Ardana-Lamas}},
  \bibinfo{author}{\bibfnamefont{I.}~\bibnamefont{Seidu}},
  \bibinfo{author}{\bibfnamefont{S.~P.} \bibnamefont{Neville}},
  \bibinfo{author}{\bibfnamefont{J.}~\bibnamefont{van~der Veen}},
  \bibinfo{author}{\bibfnamefont{V.~U.} \bibnamefont{Lanfaloni}},
  \bibinfo{author}{\bibfnamefont{M.~S.} \bibnamefont{Schuurman}},
  \bibnamefont{and} \bibinfo{author}{\bibfnamefont{H.~J.}
  \bibnamefont{W{\"o}rner}}, \bibinfo{journal}{Science}
  \textbf{\bibinfo{volume}{371}}, \bibinfo{pages}{489} (\bibinfo{year}{2021}),
  ISSN \bibinfo{issn}{0036-8075},
  \eprint{https://science.sciencemag.org/content/371/6528/489.full.pdf},
  \urlprefix\url{https://science.sciencemag.org/content/371/6528/489}.

\bibitem[{\citenamefont{Wu et~al.}(2011)\citenamefont{Wu, Boguslavskiy, Schalk,
  Schuurman, and Stolow}}]{Wu_2011}
\bibinfo{author}{\bibfnamefont{G.}~\bibnamefont{Wu}},
  \bibinfo{author}{\bibfnamefont{A.~E.} \bibnamefont{Boguslavskiy}},
  \bibinfo{author}{\bibfnamefont{O.}~\bibnamefont{Schalk}},
  \bibinfo{author}{\bibfnamefont{M.~S.} \bibnamefont{Schuurman}},
  \bibnamefont{and} \bibinfo{author}{\bibfnamefont{A.}~\bibnamefont{Stolow}},
  \bibinfo{journal}{The Journal of Chemical Physics}
  \textbf{\bibinfo{volume}{135}}, \bibinfo{pages}{164309}
  (\bibinfo{year}{2011}), \eprint{https://doi.org/10.1063/1.3652966},
  \urlprefix\url{https://doi.org/10.1063/1.3652966}.

\bibitem[{\citenamefont{MacDonell et~al.}(2016)\citenamefont{MacDonell, Schalk,
  Geng, Thomas, Feifel, Hansson, and Schuurman}}]{MacDonell_2016}
\bibinfo{author}{\bibfnamefont{R.~J.} \bibnamefont{MacDonell}},
  \bibinfo{author}{\bibfnamefont{O.}~\bibnamefont{Schalk}},
  \bibinfo{author}{\bibfnamefont{T.}~\bibnamefont{Geng}},
  \bibinfo{author}{\bibfnamefont{R.~D.} \bibnamefont{Thomas}},
  \bibinfo{author}{\bibfnamefont{R.}~\bibnamefont{Feifel}},
  \bibinfo{author}{\bibfnamefont{T.}~\bibnamefont{Hansson}}, \bibnamefont{and}
  \bibinfo{author}{\bibfnamefont{M.~S.} \bibnamefont{Schuurman}},
  \bibinfo{journal}{The Journal of Chemical Physics}
  \textbf{\bibinfo{volume}{145}}, \bibinfo{pages}{114306}
  (\bibinfo{year}{2016}), \eprint{https://doi.org/10.1063/1.4962170},
  \urlprefix\url{https://doi.org/10.1063/1.4962170}.

\bibitem[{\citenamefont{MacDonell and
  Schuurman}(2018)}]{MacDonell_ChemPhys_2018}
\bibinfo{author}{\bibfnamefont{R.~J.} \bibnamefont{MacDonell}}
  \bibnamefont{and} \bibinfo{author}{\bibfnamefont{M.~S.}
  \bibnamefont{Schuurman}}, \bibinfo{journal}{Chem. Phys.}
  \textbf{\bibinfo{volume}{515}}, \bibinfo{pages}{360} (\bibinfo{year}{2018}),
  ISSN \bibinfo{issn}{0301-0104}, \bibinfo{note}{ultrafast Photoinduced
  Processes in Polyatomic Molecules:Electronic Structure, Dynamics and
  Spectroscopy (Dedicated to Wolfgang Domcke on the occasion of his 70th
  birthday)},
  \urlprefix\url{http://www.sciencedirect.com/science/article/pii/S0301010418305871}.

\bibitem[{\citenamefont{Herperger et~al.}(2020)\citenamefont{Herperger,
  R{\"o}der, MacDonell, Boguslavskiy, Skov, Stolow, and
  Schuurman}}]{Herperger_JChemPhys_2020}
\bibinfo{author}{\bibfnamefont{K.~R.} \bibnamefont{Herperger}},
  \bibinfo{author}{\bibfnamefont{A.}~\bibnamefont{R{\"o}der}},
  \bibinfo{author}{\bibfnamefont{R.~J.} \bibnamefont{MacDonell}},
  \bibinfo{author}{\bibfnamefont{A.~E.} \bibnamefont{Boguslavskiy}},
  \bibinfo{author}{\bibfnamefont{A.~B.} \bibnamefont{Skov}},
  \bibinfo{author}{\bibfnamefont{A.}~\bibnamefont{Stolow}}, \bibnamefont{and}
  \bibinfo{author}{\bibfnamefont{M.~S.} \bibnamefont{Schuurman}},
  \bibinfo{journal}{J. Chem. Phys.} \textbf{\bibinfo{volume}{153}},
  \bibinfo{pages}{244307} (\bibinfo{year}{2020}),
  \eprint{https://doi.org/10.1063/5.0031689},
  \urlprefix\url{https://doi.org/10.1063/5.0031689}.

\bibitem[{\citenamefont{Levine and
  Mart{\'i}nez}(2009)}]{Levine_JPhysChemA_2009}
\bibinfo{author}{\bibfnamefont{B.~G.} \bibnamefont{Levine}} \bibnamefont{and}
  \bibinfo{author}{\bibfnamefont{T.~J.} \bibnamefont{Mart{\'i}nez}},
  \bibinfo{journal}{J. Phys. Chem. A} \textbf{\bibinfo{volume}{113}},
  \bibinfo{pages}{12815} (\bibinfo{year}{2009}), \bibinfo{note}{pMID:
  19813720}, \eprint{https://doi.org/10.1021/jp907111u},
  \urlprefix\url{https://doi.org/10.1021/jp907111u}.

\bibitem[{\citenamefont{Ben-Nun and
  Mart{\'i}nez}(1998)}]{BenNum_ChemPhysLett_1998}
\bibinfo{author}{\bibfnamefont{M.}~\bibnamefont{Ben-Nun}} \bibnamefont{and}
  \bibinfo{author}{\bibfnamefont{T.~J.} \bibnamefont{Mart{\'i}nez}},
  \bibinfo{journal}{Chem. Phys. Lett.} \textbf{\bibinfo{volume}{298}},
  \bibinfo{pages}{57} (\bibinfo{year}{1998}), ISSN \bibinfo{issn}{0009-2614},
  \urlprefix\url{http://www.sciencedirect.com/science/article/pii/S0009261498011154}.

\bibitem[{\citenamefont{Mori et~al.}(2012)\citenamefont{Mori, Glover,
  Schuurman, and Mart{\'i}nez}}]{Mori_JPhysChemA_2012}
\bibinfo{author}{\bibfnamefont{T.}~\bibnamefont{Mori}},
  \bibinfo{author}{\bibfnamefont{W.~J.} \bibnamefont{Glover}},
  \bibinfo{author}{\bibfnamefont{M.~S.} \bibnamefont{Schuurman}},
  \bibnamefont{and} \bibinfo{author}{\bibfnamefont{T.~J.}
  \bibnamefont{Mart{\'i}nez}}, \bibinfo{journal}{J. Phys. Chem. A}
  \textbf{\bibinfo{volume}{116}}, \bibinfo{pages}{2808} (\bibinfo{year}{2012}),
  \bibinfo{note}{pMID: 22148837}, \eprint{https://doi.org/10.1021/jp2097185},
  \urlprefix\url{https://doi.org/10.1021/jp2097185}.

\bibitem[{\citenamefont{MacDonell et~al.}(2020)\citenamefont{MacDonell,
  Corrales, Boguslavskiy, Bañares, Stolow, and
  Schuurman}}]{MacDonell_JChemPhys_2020}
\bibinfo{author}{\bibfnamefont{R.~J.} \bibnamefont{MacDonell}},
  \bibinfo{author}{\bibfnamefont{M.~E.} \bibnamefont{Corrales}},
  \bibinfo{author}{\bibfnamefont{A.~E.} \bibnamefont{Boguslavskiy}},
  \bibinfo{author}{\bibfnamefont{L.}~\bibnamefont{Bañares}},
  \bibinfo{author}{\bibfnamefont{A.}~\bibnamefont{Stolow}}, \bibnamefont{and}
  \bibinfo{author}{\bibfnamefont{M.~S.} \bibnamefont{Schuurman}},
  \bibinfo{journal}{J. Chem. Phys.} \textbf{\bibinfo{volume}{152}},
  \bibinfo{pages}{084308} (\bibinfo{year}{2020}),
  \eprint{https://doi.org/10.1063/1.5139446},
  \urlprefix\url{https://doi.org/10.1063/1.5139446}.

\bibitem[{\citenamefont{Ben-Nun et~al.}(2000)\citenamefont{Ben-Nun,
  Quenneville, and Mart{\'i}nez}}]{BenNun_JPhysChemA_2000}
\bibinfo{author}{\bibfnamefont{M.}~\bibnamefont{Ben-Nun}},
  \bibinfo{author}{\bibfnamefont{J.}~\bibnamefont{Quenneville}},
  \bibnamefont{and} \bibinfo{author}{\bibfnamefont{T.~J.}
  \bibnamefont{Mart{\'i}nez}}, \bibinfo{journal}{J. Phys. Chem. A}
  \textbf{\bibinfo{volume}{104}}, \bibinfo{pages}{5161} (\bibinfo{year}{2000}),
  \eprint{https://doi.org/10.1021/jp994174i},
  \urlprefix\url{https://doi.org/10.1021/JP994174}.

\bibitem[{\citenamefont{Mart{\'i}nez}(2006)}]{Martinez_AccChemRes_2006}
\bibinfo{author}{\bibfnamefont{T.~J.} \bibnamefont{Mart{\'i}nez}},
  \bibinfo{journal}{Acc. Chem. Res.} \textbf{\bibinfo{volume}{39}},
  \bibinfo{pages}{119} (\bibinfo{year}{2006}), \bibinfo{note}{pMID: 16489731},
  \eprint{https://doi.org/10.1021/ar040202q},
  \urlprefix\url{https://doi.org/10.1021/ar040202q}.

\bibitem[{\citenamefont{Mignolet and Curchod}(2018)}]{churchod_2018}
\bibinfo{author}{\bibfnamefont{B.}~\bibnamefont{Mignolet}} \bibnamefont{and}
  \bibinfo{author}{\bibfnamefont{B.~F.~E.} \bibnamefont{Curchod}},
  \bibinfo{journal}{J. Chem. Phys.} \textbf{\bibinfo{volume}{148}},
  \bibinfo{pages}{134110} (\bibinfo{year}{2018}),
  \eprint{https://doi.org/10.1063/1.5022877},
  \urlprefix\url{https://doi.org/10.1063/1.5022877}.

\bibitem[{\citenamefont{Yang et~al.}(2009)\citenamefont{Yang, Coe, Kaduk, and
  Mart{\'i}nez}}]{Yang_JChemPhys_2009}
\bibinfo{author}{\bibfnamefont{S.}~\bibnamefont{Yang}},
  \bibinfo{author}{\bibfnamefont{J.~D.} \bibnamefont{Coe}},
  \bibinfo{author}{\bibfnamefont{B.}~\bibnamefont{Kaduk}}, \bibnamefont{and}
  \bibinfo{author}{\bibfnamefont{T.~J.} \bibnamefont{Mart{\'i}nez}},
  \bibinfo{journal}{J. Chem. Phys.} \textbf{\bibinfo{volume}{130}},
  \bibinfo{pages}{134113} (\bibinfo{year}{2009}),
  \eprint{https://doi.org/10.1063/1.3103930},
  \urlprefix\url{https://doi.org/10.1063/1.3103930}.

\bibitem[{\citenamefont{Grimme}(1996)}]{Grimme_DFTCIS_ChemPhysLett_1996}
\bibinfo{author}{\bibfnamefont{S.}~\bibnamefont{Grimme}},
  \bibinfo{journal}{Chem. Phys. Lett.} \textbf{\bibinfo{volume}{259}},
  \bibinfo{pages}{128} (\bibinfo{year}{1996}), ISSN \bibinfo{issn}{0009-2614},
  \urlprefix\url{http://www.sciencedirect.com/science/article/pii/0009261496007221}.

\bibitem[{\citenamefont{Grimme and Waletzke}(1999)}]{Grimme_DFTMRCI_JCP_1999}
\bibinfo{author}{\bibfnamefont{S.}~\bibnamefont{Grimme}} \bibnamefont{and}
  \bibinfo{author}{\bibfnamefont{M.}~\bibnamefont{Waletzke}},
  \bibinfo{journal}{J. Chem. Phys.} \textbf{\bibinfo{volume}{111}},
  \bibinfo{pages}{5645} (\bibinfo{year}{1999}),
  \eprint{https://doi.org/10.1063/1.479866},
  \urlprefix\url{https://doi.org/10.1063/1.479866}.

\bibitem[{\citenamefont{Heil et~al.}(2016)\citenamefont{Heil, Gollnisch,
  Kleinschmidt, and Marian}}]{Heil_DFTMRCI_MolPhys_2016}
\bibinfo{author}{\bibfnamefont{A.}~\bibnamefont{Heil}},
  \bibinfo{author}{\bibfnamefont{K.}~\bibnamefont{Gollnisch}},
  \bibinfo{author}{\bibfnamefont{M.}~\bibnamefont{Kleinschmidt}},
  \bibnamefont{and} \bibinfo{author}{\bibfnamefont{C.~M.}
  \bibnamefont{Marian}}, \bibinfo{journal}{Mol. Phys.}
  \textbf{\bibinfo{volume}{114}}, \bibinfo{pages}{407} (\bibinfo{year}{2016}),
  \eprint{https://doi.org/10.1080/00268976.2015.1076902},
  \urlprefix\url{https://doi.org/10.1080/00268976.2015.1076902}.

\bibitem[{\citenamefont{Heil and Marian}(2017)}]{Heil_DFTMRCI_JCP_2017}
\bibinfo{author}{\bibfnamefont{A.}~\bibnamefont{Heil}} \bibnamefont{and}
  \bibinfo{author}{\bibfnamefont{C.~M.} \bibnamefont{Marian}},
  \bibinfo{journal}{J. Chem. Phys.} \textbf{\bibinfo{volume}{147}},
  \bibinfo{pages}{194104} (\bibinfo{year}{2017}),
  \eprint{https://doi.org/10.1063/1.5003246},
  \urlprefix\url{https://doi.org/10.1063/1.5003246}.

\bibitem[{\citenamefont{Lyskov et~al.}(2016)\citenamefont{Lyskov, Kleinschmidt,
  and Marian}}]{Lyskov_DFTMRCI_JCP_2016}
\bibinfo{author}{\bibfnamefont{I.}~\bibnamefont{Lyskov}},
  \bibinfo{author}{\bibfnamefont{M.}~\bibnamefont{Kleinschmidt}},
  \bibnamefont{and} \bibinfo{author}{\bibfnamefont{C.~M.}
  \bibnamefont{Marian}}, \bibinfo{journal}{J. Chem. Phys.}
  \textbf{\bibinfo{volume}{144}}, \bibinfo{pages}{034104}
  (\bibinfo{year}{2016}), \eprint{https://doi.org/10.1063/1.4940036},
  \urlprefix\url{https://doi.org/10.1063/1.4940036}.

\bibitem[{\citenamefont{Kleinschmidt et~al.}(2002)\citenamefont{Kleinschmidt,
  Tatchen, and Marian}}]{Kleinschmidt_DFTMRCI_JCompChem_2002}
\bibinfo{author}{\bibfnamefont{M.}~\bibnamefont{Kleinschmidt}},
  \bibinfo{author}{\bibfnamefont{J.}~\bibnamefont{Tatchen}}, \bibnamefont{and}
  \bibinfo{author}{\bibfnamefont{C.~M.} \bibnamefont{Marian}},
  \bibinfo{journal}{J. Comput. Chem.} \textbf{\bibinfo{volume}{23}},
  \bibinfo{pages}{824} (\bibinfo{year}{2002}),
  \eprint{https://onlinelibrary.wiley.com/doi/pdf/10.1002/jcc.10064},
  \urlprefix\url{https://onlinelibrary.wiley.com/doi/abs/10.1002/jcc.10064}.

\bibitem[{\citenamefont{Marian et~al.}(2019)\citenamefont{Marian, Heil, and
  Kleinschmidt}}]{Marian_2019}
\bibinfo{author}{\bibfnamefont{C.~M.} \bibnamefont{Marian}},
  \bibinfo{author}{\bibfnamefont{A.}~\bibnamefont{Heil}}, \bibnamefont{and}
  \bibinfo{author}{\bibfnamefont{M.}~\bibnamefont{Kleinschmidt}},
  \bibinfo{journal}{Wiley Interdiscip. Rev. Comput. Mol. Sci.}
  \textbf{\bibinfo{volume}{9}}, \bibinfo{pages}{e1394} (\bibinfo{year}{2019}),
  \eprint{https://onlinelibrary.wiley.com/doi/pdf/10.1002/wcms.1394},
  \urlprefix\url{https://onlinelibrary.wiley.com/doi/abs/10.1002/wcms.1394}.

\bibitem[{\citenamefont{Seidu et~al.}(2019)\citenamefont{Seidu, Neville,
  Kleinschmidt, Heil, Marian, and Schuurman}}]{Seidu_JChemPhys_2019}
\bibinfo{author}{\bibfnamefont{I.}~\bibnamefont{Seidu}},
  \bibinfo{author}{\bibfnamefont{S.~P.} \bibnamefont{Neville}},
  \bibinfo{author}{\bibfnamefont{M.}~\bibnamefont{Kleinschmidt}},
  \bibinfo{author}{\bibfnamefont{A.}~\bibnamefont{Heil}},
  \bibinfo{author}{\bibfnamefont{C.~M.} \bibnamefont{Marian}},
  \bibnamefont{and} \bibinfo{author}{\bibfnamefont{M.~S.}
  \bibnamefont{Schuurman}}, \bibinfo{journal}{J. Chem. Phys.}
  \textbf{\bibinfo{volume}{151}}, \bibinfo{pages}{144104}
  (\bibinfo{year}{2019}).

\bibitem[{\citenamefont{Lischka et~al.}(2011)\citenamefont{Lischka, M{\"u}ller,
  Szalay, Shavitt, Pitzer, and Shepard}}]{Lischka_WIREs_2011}
\bibinfo{author}{\bibfnamefont{H.}~\bibnamefont{Lischka}},
  \bibinfo{author}{\bibfnamefont{T.}~\bibnamefont{M{\"u}ller}},
  \bibinfo{author}{\bibfnamefont{P.~G.} \bibnamefont{Szalay}},
  \bibinfo{author}{\bibfnamefont{I.}~\bibnamefont{Shavitt}},
  \bibinfo{author}{\bibfnamefont{R.~M.} \bibnamefont{Pitzer}},
  \bibnamefont{and} \bibinfo{author}{\bibfnamefont{R.}~\bibnamefont{Shepard}},
  \bibinfo{journal}{WIREs} \textbf{\bibinfo{volume}{1}}, \bibinfo{pages}{191}
  (\bibinfo{year}{2011}),
  \eprint{https://onlinelibrary.wiley.com/doi/pdf/10.1002/wcms.25},
  \urlprefix\url{https://onlinelibrary.wiley.com/doi/abs/10.1002/wcms.25}.

\bibitem[{\citenamefont{Lischka et~al.}(2012)\citenamefont{Lischka, Shepard,
  Shavitt, Pitzer, Dallos, M{\"u}ller, Szalay, Brown, Ahlrichs, B{\"o}hm
  et~al.}}]{Lischka_COLUMBUS_2015}
\bibinfo{author}{\bibfnamefont{H.}~\bibnamefont{Lischka}},
  \bibinfo{author}{\bibfnamefont{R.}~\bibnamefont{Shepard}},
  \bibinfo{author}{\bibfnamefont{I.}~\bibnamefont{Shavitt}},
  \bibinfo{author}{\bibfnamefont{R.~M.} \bibnamefont{Pitzer}},
  \bibinfo{author}{\bibfnamefont{M.}~\bibnamefont{Dallos}},
  \bibinfo{author}{\bibfnamefont{T.}~\bibnamefont{M{\"u}ller}},
  \bibinfo{author}{\bibfnamefont{P.~G.} \bibnamefont{Szalay}},
  \bibinfo{author}{\bibfnamefont{F.~B.} \bibnamefont{Brown}},
  \bibinfo{author}{\bibfnamefont{R.}~\bibnamefont{Ahlrichs}},
  \bibinfo{author}{\bibfnamefont{H.~J.} \bibnamefont{B{\"o}hm}},
  \bibnamefont{et~al.}, \bibinfo{journal}{COLUMBUS, an {\it ab initio}
  electronic structure program, release 7.0}  (\bibinfo{year}{2012}).

\bibitem[{\citenamefont{Coe et~al.}(2007)\citenamefont{Coe, Levine, and
  Mart{\'i}nez}}]{Coe_2007}
\bibinfo{author}{\bibfnamefont{J.~D.} \bibnamefont{Coe}},
  \bibinfo{author}{\bibfnamefont{B.~G.} \bibnamefont{Levine}},
  \bibnamefont{and} \bibinfo{author}{\bibfnamefont{T.~J.}
  \bibnamefont{Mart{\'i}nez}}, \bibinfo{journal}{J. Phys. Chem. A}
  \textbf{\bibinfo{volume}{111}}, \bibinfo{pages}{11302}
  (\bibinfo{year}{2007}), \bibinfo{note}{pMID: 17602455},
  \eprint{https://doi.org/10.1021/jp072027b},
  \urlprefix\url{https://doi.org/10.1021/jp072027b}.

\bibitem[{tur(2007)}]{turbomole_v6.1}
\emph{\bibinfo{title}{Turbomole v6.1 2009, a development of university of
  karlsruhe and forschungszentrum karlsruhe gmbh, turbomole gmbh}}
  (\bibinfo{year}{2007}), \urlprefix\url{http://www.turbomole.com}.

\bibitem[{\citenamefont{Aidas et~al.}(2014)\citenamefont{Aidas, Angeli, Bak,
  Bakken, Bast, Boman, Christiansen, Cimiraglia, Coriani, Dahle
  et~al.}}]{daltonpaper}
\bibinfo{author}{\bibfnamefont{K.}~\bibnamefont{Aidas}},
  \bibinfo{author}{\bibfnamefont{C.}~\bibnamefont{Angeli}},
  \bibinfo{author}{\bibfnamefont{K.~L.} \bibnamefont{Bak}},
  \bibinfo{author}{\bibfnamefont{V.}~\bibnamefont{Bakken}},
  \bibinfo{author}{\bibfnamefont{R.}~\bibnamefont{Bast}},
  \bibinfo{author}{\bibfnamefont{L.}~\bibnamefont{Boman}},
  \bibinfo{author}{\bibfnamefont{O.}~\bibnamefont{Christiansen}},
  \bibinfo{author}{\bibfnamefont{R.}~\bibnamefont{Cimiraglia}},
  \bibinfo{author}{\bibfnamefont{S.}~\bibnamefont{Coriani}},
  \bibinfo{author}{\bibfnamefont{P.}~\bibnamefont{Dahle}},
  \bibnamefont{et~al.}, \bibinfo{journal}{WIREs Comput.~Mol.~Sci.}
  \textbf{\bibinfo{volume}{4}}, \bibinfo{pages}{269} (\bibinfo{year}{2014}).

\bibitem[{\citenamefont{Becke}(1988)}]{Becke_B3LYP_PhysRevA_1988}
\bibinfo{author}{\bibfnamefont{A.~D.} \bibnamefont{Becke}},
  \bibinfo{journal}{Phys. Rev. A} \textbf{\bibinfo{volume}{38}},
  \bibinfo{pages}{3098} (\bibinfo{year}{1988}),
  \urlprefix\url{https://link.aps.org/doi/10.1103/PhysRevA.38.3098}.

\bibitem[{\citenamefont{Becke}(1993{\natexlab{a}})}]{Becke_B3LYP_JCP_1993}
\bibinfo{author}{\bibfnamefont{A.~D.} \bibnamefont{Becke}},
  \bibinfo{journal}{J. Chem. Phys.} \textbf{\bibinfo{volume}{98}},
  \bibinfo{pages}{1372} (\bibinfo{year}{1993}{\natexlab{a}}),
  \eprint{https://doi.org/10.1063/1.464304},
  \urlprefix\url{https://doi.org/10.1063/1.464304}.

\bibitem[{\citenamefont{Becke}(1993{\natexlab{b}})}]{Becke_JChemPhys_1993}
\bibinfo{author}{\bibfnamefont{A.~D.} \bibnamefont{Becke}},
  \bibinfo{journal}{J. Chem. Phys.} \textbf{\bibinfo{volume}{98}},
  \bibinfo{pages}{1372} (\bibinfo{year}{1993}{\natexlab{b}}),
  \eprint{https://doi.org/10.1063/1.464304},
  \urlprefix\url{https://doi.org/10.1063/1.464304}.

\bibitem[{\citenamefont{Kabsch}(1976)}]{Kabsch_ActaCryst_1976}
\bibinfo{author}{\bibfnamefont{W.}~\bibnamefont{Kabsch}},
  \bibinfo{journal}{Acta Cryst.} \textbf{\bibinfo{volume}{32}},
  \bibinfo{pages}{922} (\bibinfo{year}{1976}),
  \urlprefix\url{https://doi.org/10.1107/S0567739476001873}.

\bibitem[{\citenamefont{Neville
  et~al.}(2018{\natexlab{b}})\citenamefont{Neville, Chergui, Stolow, and
  Schuurman}}]{Neville_ADC_PhysRevLett_2018}
\bibinfo{author}{\bibfnamefont{S.~P.} \bibnamefont{Neville}},
  \bibinfo{author}{\bibfnamefont{M.}~\bibnamefont{Chergui}},
  \bibinfo{author}{\bibfnamefont{A.}~\bibnamefont{Stolow}}, \bibnamefont{and}
  \bibinfo{author}{\bibfnamefont{M.~S.} \bibnamefont{Schuurman}},
  \bibinfo{journal}{Phys. Rev. Lett.} \textbf{\bibinfo{volume}{120}},
  \bibinfo{pages}{243001} (\bibinfo{year}{2018}{\natexlab{b}}),
  \urlprefix\url{https://link.aps.org/doi/10.1103/PhysRevLett.120.243001}.

\end{thebibliography}


\begin{thebibliography}{1}
\expandafter\ifx\csname natexlab\endcsname\relax\def\natexlab#1{#1}\fi
\expandafter\ifx\csname bibnamefont\endcsname\relax
  \def\bibnamefont#1{#1}\fi
\expandafter\ifx\csname bibfnamefont\endcsname\relax
  \def\bibfnamefont#1{#1}\fi
\expandafter\ifx\csname citenamefont\endcsname\relax
  \def\citenamefont#1{#1}\fi
\expandafter\ifx\csname url\endcsname\relax
  \def\url#1{\texttt{#1}}\fi
\expandafter\ifx\csname urlprefix\endcsname\relax\def\urlprefix{URL }\fi
\providecommand{\bibinfo}[2]{#2}
\providecommand{\eprint}[2][]{\url{#2}}

\bibitem[{\citenamefont{McLaren et~al.}(1987)\citenamefont{McLaren, Clark,
  Ishii, and Hitchcock}}]{ethylene}
\bibinfo{author}{\bibfnamefont{R.}~\bibnamefont{McLaren}},
  \bibinfo{author}{\bibfnamefont{S.~A.~C.} \bibnamefont{Clark}},
  \bibinfo{author}{\bibfnamefont{I.}~\bibnamefont{Ishii}}, \bibnamefont{and}
  \bibinfo{author}{\bibfnamefont{A.~P.} \bibnamefont{Hitchcock}},
  \bibinfo{journal}{Phys. Rev. A} \textbf{\bibinfo{volume}{36}},
  \bibinfo{pages}{1683} (\bibinfo{year}{1987}),
  \urlprefix\url{https://link.aps.org/doi/10.1103/PhysRevA.36.1683}.

\end{thebibliography}

\end{document}


\title{Supplementary Information - Resolving Competing Conical Intersection Pathways: Time-Resolved X-ray Absorption Spectroscopy of trans-1,3-Butadiene} 

\author{Issaka Seidu}
\affiliation{National Research Council of Canada, 100 Sussex Drive,
  Ottawa, Ontario K1A 0R6, Canada}

\author{Simon P. Neville}
\affiliation{National Research Council of Canada, 100 Sussex Drive,
  Ottawa, Ontario K1A 0R6, Canada}

\author{Ryan J. MacDonell}
\affiliation{University of Sydney, Sydney, Australia}

\author{Michael S. Schuurman}
\affiliation{National Research Council of Canada, 100 Sussex Drive,
  Ottawa, Ontario K1A 0R6, Canada}
\affiliation{Department of Chemistry and Biomolecular Sciences,
  University of Ottawa, 10 Marie Curie, Ottawa, Ontario, K1N 6N5,
  Canada}

\maketitle

\begin{figure}[hbt!]
    \centering
    \includegraphics[width=1.\textwidth]{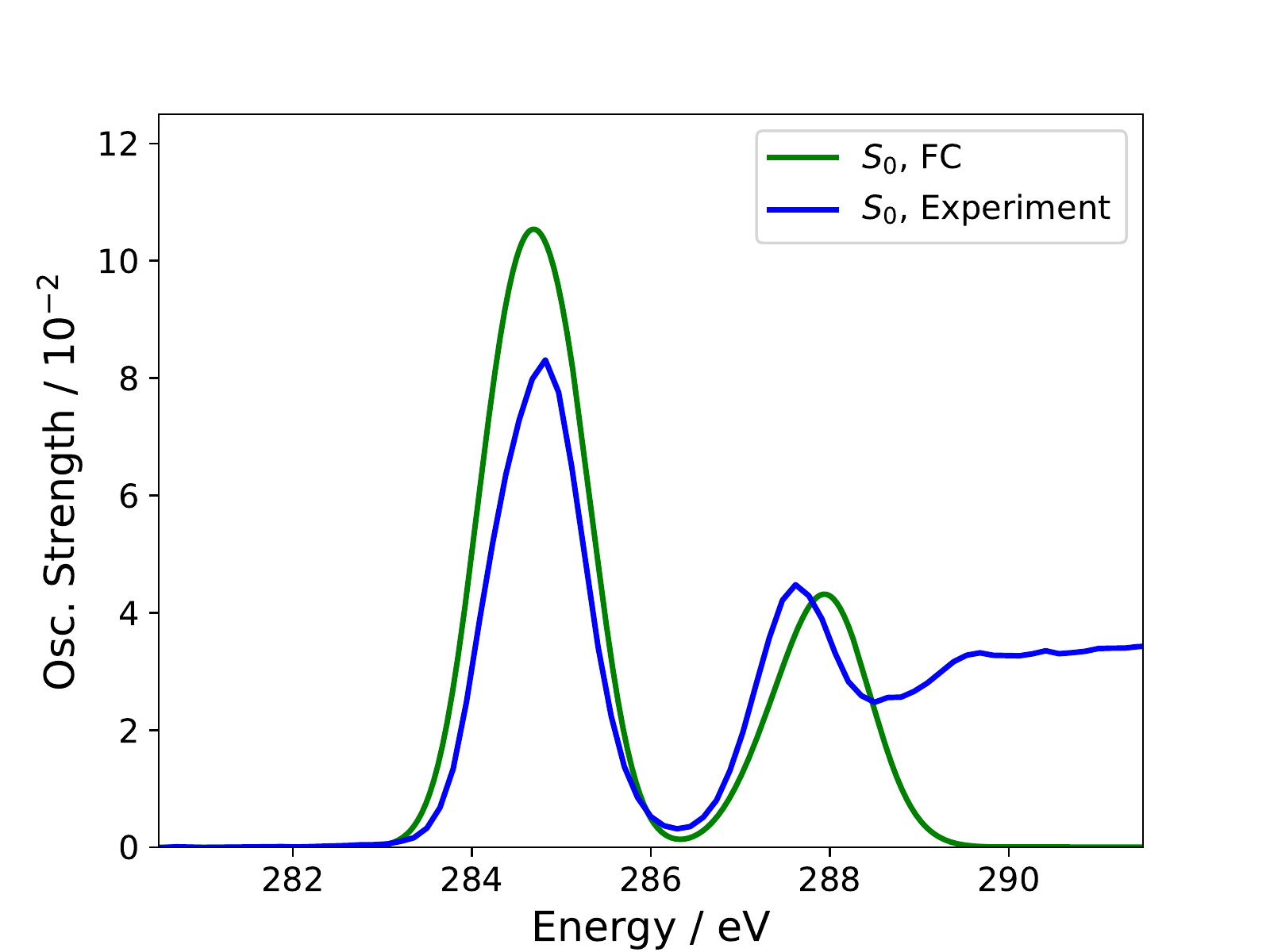}
    \caption{Calculated XAS at CVS-DFT/MRCI/aug-cc-pVDZ level of theory for $S_0$ state at the FC compared to the experimental EELS spectrum\cite{ethylene}. The calculated spectrum is convoluted with a FWHM = 1.0 eV and shifted by 3.5 eV to higher energy to facilitate comparison to experimental spectrum.}
    \label{fig:fc_expt_xas}
\end{figure}

\begin{table}[htb!]
  \caption{Transoid NTOs of the ground-state and excited-state XAS.}
  \begin{center}
    \includegraphics[width=.9\textwidth]{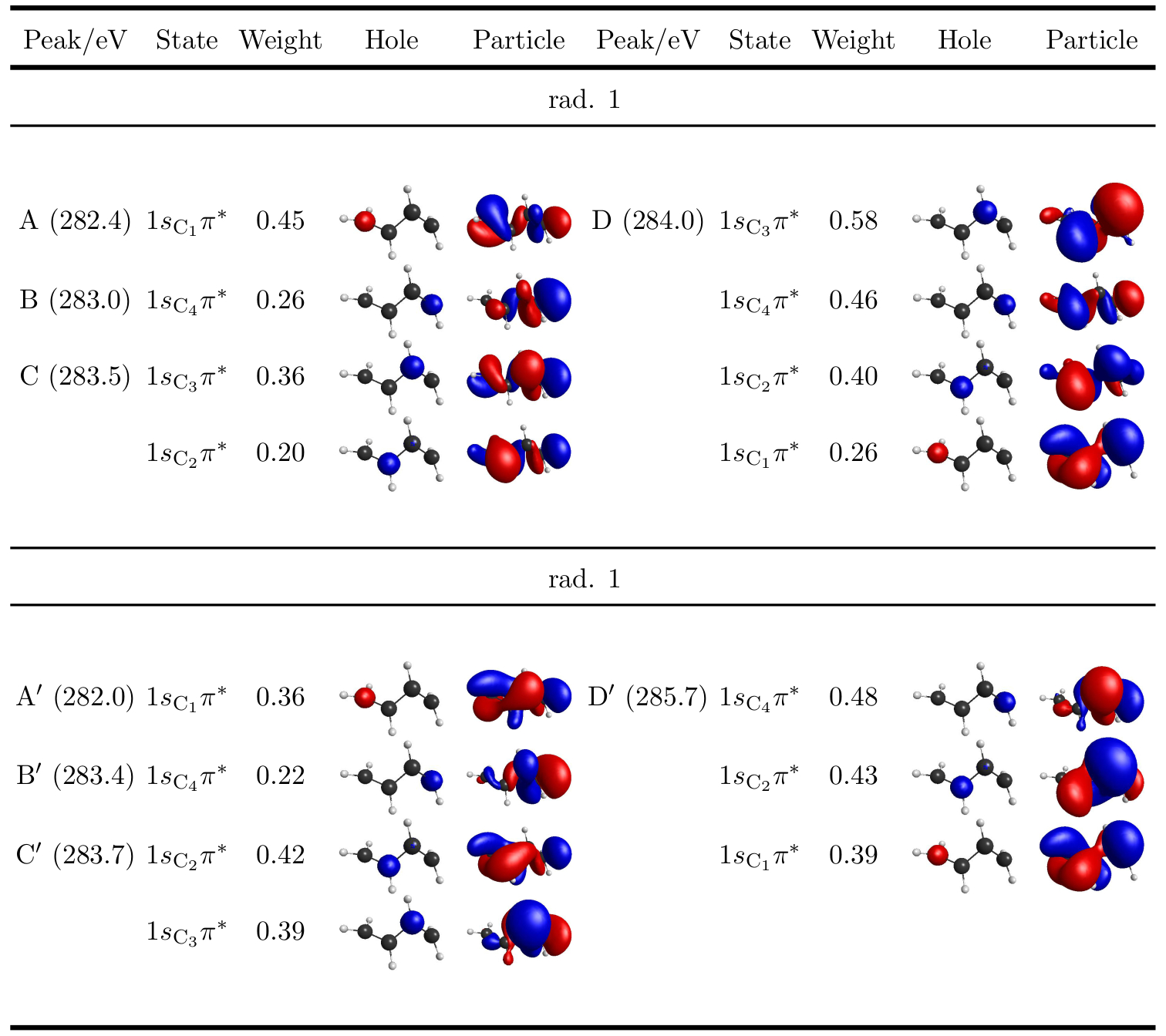}
  \end{center}
  \label{tab:transoid_nto}
\end{table}

\begin{table}[htb!]
  \caption{Tw-Py NTOs of the ground-state and excited-state XAS.}
  \begin{center}
    \includegraphics[width=.9\textwidth]{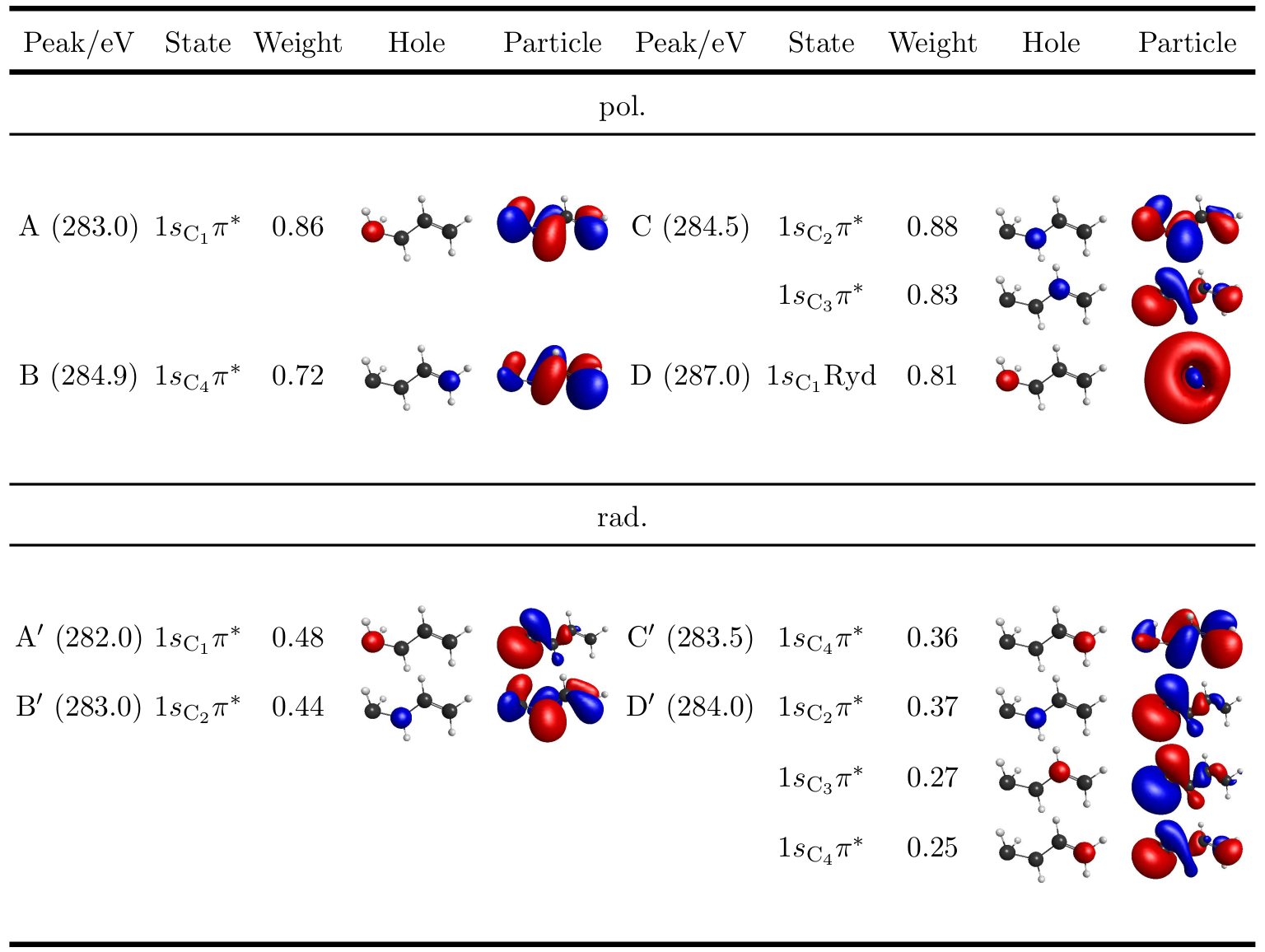}
  \end{center}
  \label{tab:twpy_nto}
\end{table}

\begin{figure}[hbt!]
    \centering
    \includegraphics[width=1.\textwidth]{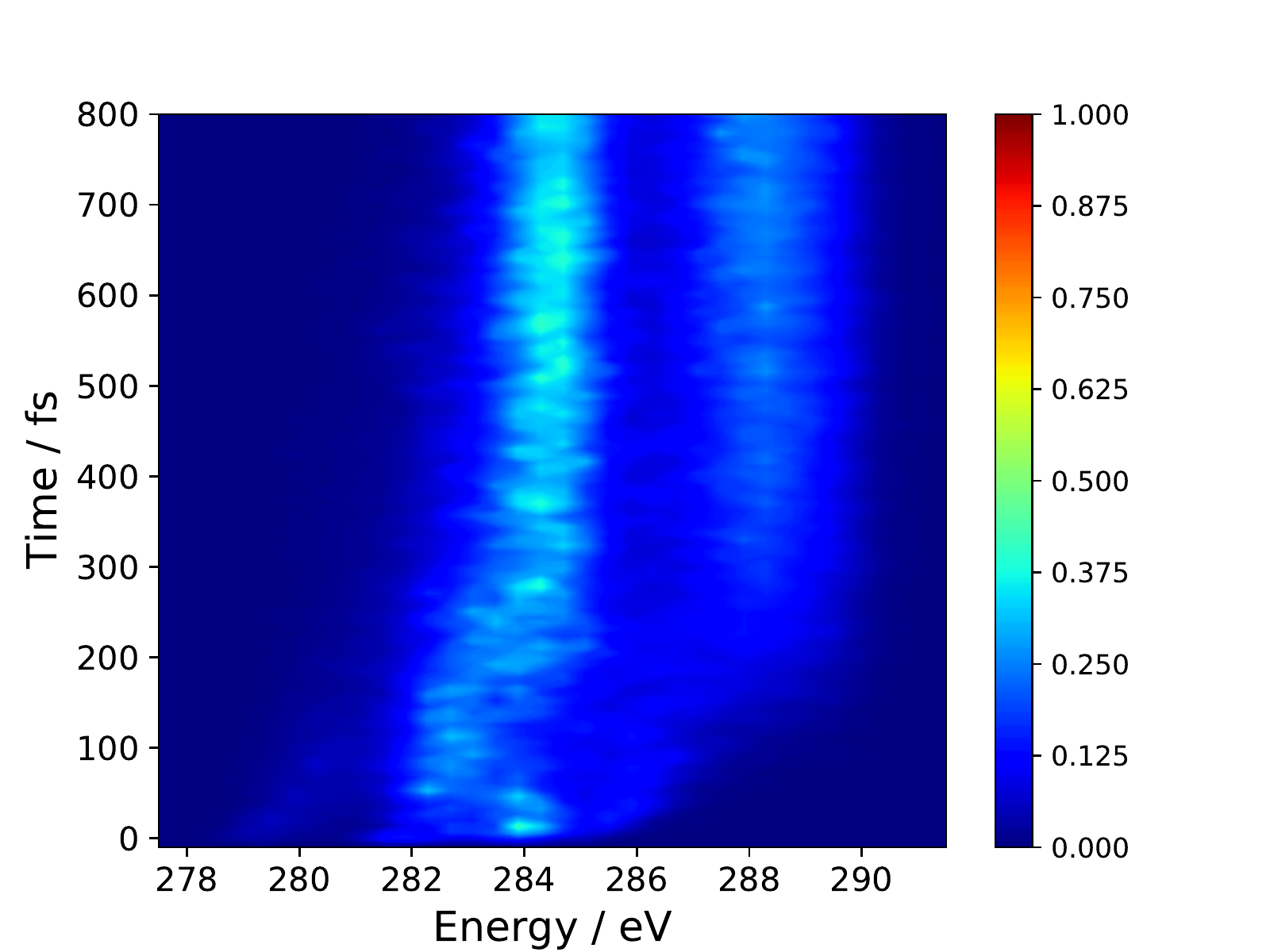}
    \caption{Simulated parallel polarization TRXAS spectrum employing the results of AIMS dynamics simulations and CVS-DFT/MRCI X-ray absorption cross-sections. An isotropic axis distribution was assumed.}
    \label{fig:trxas_para}
\end{figure}

\begin{figure}[hbt!]
    \centering
    \includegraphics[width=1.\textwidth]{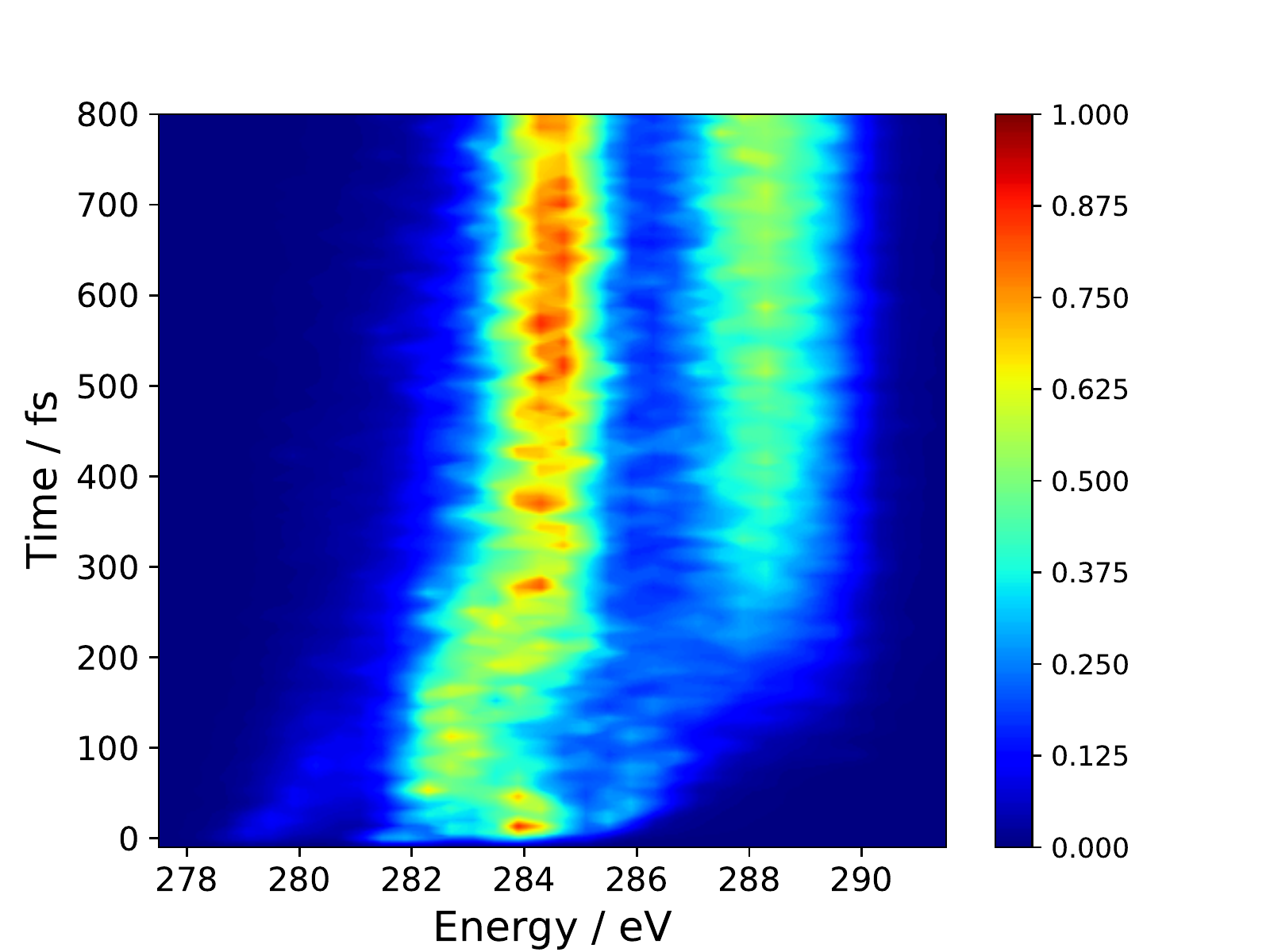}
    \caption{Simulated perpendicular polarization TRXAS spectrum employing the results of AIMS dynamics simulations and CVS-DFT/MRCI X-ray absorption cross-sections. An isotropic axis distribution was assumed.}
    \label{fig:trxas_perp}
\end{figure}

\clearpage

\begin{table}[htb!]
  \caption{MR-CIS optimized geometry of the terminal Tw-Py MECI.}
  \begin{center}
     \begin{tabular}{lrrr}
     \hline
      C     &    1.841822   &      0.081208     &     0.057792\\
      C     &    0.689252   &     -0.634435     &    -0.026686\\
      C     &   -0.601894   &      0.013296     &    -0.099864\\
      C     &   -1.816004   &     -0.716548     &     0.021817\\
      H     &    2.804940   &     -0.398401     &     0.083936\\
      H     &    1.828613   &      1.158278     &     0.096778\\
      H     &    0.686535   &     -1.710894     &    -0.041153\\
      H     &   -0.526661   &      1.104012     &    -0.129623\\
      H     &   -2.674612   &     -0.198314     &    -0.418121\\
      H     &   -1.816589   &     -0.006067     &     0.962993\\
     \hline
     \end{tabular}
  \end{center} 
  \label{tab:twpy_meci}
\end{table}

\begin{table}[htb!]
  \caption{MR-CIS optimized geometry of the medial Tw-Py (mTw-Py) MECI.}
  \begin{center}
     \begin{tabular}{lrrr}
     \hline
      C    &     1.363654    &    -1.476754    &      1.033232\\
      C    &     0.534891    &    -0.949371    &      0.036100\\
      C    &    -0.670059    &    -0.158282    &      0.076533\\
      C    &    -1.952813    &    -0.537649    &     -0.176377\\
      H    &     2.387907    &    -1.747905    &      0.815252\\
      H    &     1.012393    &    -1.807281    &      2.008700\\
      H    &     0.204967    &    -2.070707    &      0.266985\\
      H    &    -0.474465    &     0.895795    &      0.220505\\
      H    &    -2.217817    &    -1.566880    &     -0.361516\\
      H    &    -2.737362    &     0.195240    &     -0.243732\\
     \hline
     \end{tabular}
  \end{center} 
  \label{tab:mtwpy_meci}
\end{table}

\begin{table}[htb!]
  \caption{MR-CIS optimized geometry of the Transoid MECI.}
  \begin{center}
     \begin{tabular}{lrrr}
     \hline
    C  & -1.6964209992   &   0.0784200000   &  -0.3501989999\\
    C  &  1.6837099992   &  -0.1305659999   &  -0.2390789999\\
    C  & -0.4545049998   &  -0.4328799998   &   0.2369249999\\
    C  &  0.5936899998   &   0.4793759998   &   0.5337419997\\
    H  & -1.6246319993   &   0.7433189997   &  -1.2007419994\\
    H  &  2.0114239991   &  -1.1354329995   &   0.0052020000\\
    H  & -2.5595129988   &   0.2052749999   &   0.2976069999\\
    H  &  1.9171149992   &   0.2301169999   &  -1.2359359995\\
    H  & -0.2929549998   &  -1.4913829993   &   0.4101809998\\
    H  &  0.3987159998   &   1.5435609993   &   0.4642279998\\
     \hline
     \end{tabular}
  \end{center} 
  \label{tab:trans_meci}
\end{table}

\begin{table}[htb!]
  \caption{Amplitude-weighted average polarized nuclear structure.}
  \begin{center}
     \begin{tabular}{lrrr}
     \hline
     C  &   -1.74157096  &    0.29181917   &   0.03320491\\
     C  &   -0.59423943  &   -0.41124976   &   0.05067867\\
     C  &    0.65180151  &    0.27447095   &   0.12570390\\
     C  &    1.75790994  &   -0.27729612   &  -0.03407427\\
     H  &   -1.90570291  &    0.05433743   &  -0.93474674\\
     H  &   -2.40665996  &    0.11383504   &   0.51013051\\
     H  &   -0.42502721  &   -1.47340208   &  -0.13745125\\
     H  &    0.52785253  &    1.31803074   &   0.23989687\\
     H  &    2.58588754  &    0.22764448   &   0.04246069\\
     H  &    1.78407045  &   -1.29318991   &  -0.21697182\\
     \hline
     \end{tabular}
  \end{center} 
  \label{tab:avtwpy_meci}
\end{table}

\begin{table}[htb!]
  \caption{Amplitude-weighted average radicaloid nuclear structure.}
  \begin{center}
     \begin{tabular}{lrrr}
     \hline
     C   &   1.37910228   &  -0.53731288   &   0.97351603\\
     C   &   0.40371732   &  -0.43791040   &  -0.14218166\\
     C   &  -0.16702140   &   0.75629531   &  -0.50559170\\
     C   &  -1.56735250   &   0.27035629   &  -0.42560908\\
     H   &   2.29933349   &  -0.49777234   &   0.81387909\\
     H   &   1.21559161   &  -0.49793588   &   1.82447301\\
     H   &   0.20243305   &  -1.29936522   &  -0.70451660\\
     H   &   0.10246264   &   1.68714073   &  -0.17654197\\
     H   &  -1.93632648   &  -0.43172744   &  -1.06541448\\
     H   &  -2.09593225   &   0.49126615   &   0.38434477\\
     \hline
     \end{tabular}
  \end{center} 
  \label{tab:avtrans_meci}
\end{table}

\newpage
\clearpage
\bibliography{references}